\documentclass[onecolumn,prd,amssymb,amsmath,preprintnumbers,aps,nofootinbib,superscriptaddress,11pt]{revtex4-1}
\usepackage[a4paper, portrait, margin=0.7in]{geometry}
\usepackage[utf8]{inputenc}
\usepackage{multirow}
\usepackage{lineno}
\usepackage{bm,color}
\usepackage[dvipsnames]{xcolor}
\usepackage{slashed} 
\usepackage{graphicx}
\usepackage{setspace}
\usepackage{multirow,booktabs,colortbl}
\usepackage{soul} 
\usepackage{multirow}
\usepackage{comment}
\usepackage{epstopdf} 
\usepackage{mathtools}
\usepackage{draftfigure}
\usepackage{upgreek}
\usepackage{appendix}
\usepackage[colorlinks=true
,urlcolor=blue
,anchorcolor=blue
,citecolor=blue 
,filecolor=blue
,linkcolor=blue
,menucolor=blue
,linktocpage=true
,pdfproducer=medialab
,pdfa=true
]{hyperref}

\def\figureautorefname~#1\null{Fig.\,#1\null}

\def\equationautorefname~#1\null{Eq.\,(#1)\null}

\newcommand{\SM}{\rm SM}

\newcommand{\beq}{\begin{equation}}
\newcommand{\eeq}{\end{equation}}
\newcommand{\bea}{\begin{eqnarray}}
\newcommand{\eea}{\end{eqnarray}}
\newcommand{\ba}{\begin{array}}
\newcommand{\ea}{\end{array}}

\newcommand{\cmrule}{\midrule[0.25mm]}

\newcommand{\ctoprule}{\toprule[0.5mm]}
\newcommand{\cbottomrule}{\bottomrule[0.5mm]}

\def\m1{M_1}
\def\m2{M_2}
\def\m3{M_3}

\def\ch10{\tilde \chi^0_1}

\def\gev{\,{\rm GeV}}
\def\mev{\,{\rm MeV}}

\def\to{\rightarrow}

\newcommand{\lsim}{\mathrel{\mathop{\kern 0pt \rlap
  {\raise.2ex\hbox{$<$}}}
  \lower.9ex\hbox{\kern-.190em $\sim$}}}
\newcommand{\gsim}{\mathrel{\mathop{\kern 0pt \rlap
  {\raise.2ex\hbox{$>$}}}
  \lower.9ex\hbox{\kern-.190em $\sim$}}}

\def\fbi{\,{\rm fb}^{-1}}
\def\abi{\,{\rm ab}^{-1}}

\newcommand{\ifb}{{\,{\rm fb}^{-1}}}
\newcommand{\iab}{{\,{\rm ab}^{-1}}}

\begin{document}

\title{Higgs Precision at a 125 GeV Muon Collider}
\author{Jorge de Blas}
\email{deblasm@ugr.es}
\affiliation{CAFPE and Departamento de F\'isica Te\'orica y del Cosmos, Universidad de Granada, Campus de Fuentenueva, E–18071 Granada, Spain}
\author{Jiayin Gu}
\email{jiayin\_gu@fudan.edu.cn}
\affiliation{Department of Physics and Center for Field Theory and Particle Physics, Fudan University, Shanghai 200438, China}
\affiliation{Key Laboratory of Nuclear Physics and Ion-beam Application (MOE), Fudan University, Shanghai 200433, China}
\author{Zhen Liu}
\email{zliuphys@umn.edu}
\thanks{\href{https://orcid.org/0000-0002-3143-1976}{0000-0002-3143-1976}}
\affiliation{School of Physics and Astronomy, University of Minnesota, Minneapolis, MN 55455, USA}

%\date{March 2021}

\begin{abstract}
 
The $s$-channel resonant production of the Higgs boson at a 125 GeV muon collider enables a unique way to determine the Higgs properties. However, a clear picture of the achievable Higgs precision has not yet been established. 
We perform a phenomenological study of the Higgs measurements at such resonant muon collider Higgs factory and present a systematic, detailed, and consistent extraction of Higgs precision measurements. Many new aspects about the lineshape scan, including the scaling with luminosity, optimal scan range, minimal scan steps, correlations with exclusive measurement, effective cross-section modeling, etc., are quantitatively studied in this work. 
All major exclusive Higgs channels are simulated and analyzed with Standard Model background, detection efficiencies, acceptance, angular distributions, and cross-channel correlations.
Global analyses of the Higgs couplings are performed in the $\kappa$ framework and the effective-field-theory one. The results suggest that the 125 GeV muon-collider Higgs factory provides significant improvement to the Higgs coupling reach of the HL-LHC and provides independent and distinct Higgs precision information concerning future $e^+e^-$ colliders. 
We report results for both $5~\fbi$ and $20~\fbi$ integrated luminosity.
These results provide comprehensive and quantitative physics understandings helpful in planning for the muon collider roadmap and global high-energy physics programs.
\end{abstract}

\maketitle

%!!!!!!!!!!!!!!!!!!!!!!!!!!!!!!!!!!!!!!!!!!!!!!!!!!

%====================
{\small 
\tableofcontents}
%====================

\section{Introduction}

%=============================================================
Precision measurements of the Higgs properties are a powerful probe of new physics, and they play a central role in the physics programs of future colliders. Among various options, the resonant muon collider Higgs factory operating at the Higgs pole is particular in accessing Higgs precision information. The powerful resonance production and lineshape scan allow a muon collider to extract the Higgs properties differently compared to other colliders. 
However, so far, we still do not have a global picture of how well a $\mu^+\mu^-$ resonant Higgs factory can perform. 
Past developments have focused on the Higgs width precision determination, and very little is known about the overall Higgs precision results. Part of the reason is that one to two orders of magnitude fewer Higgses is expected from the muon collider. However, the determination of the Higgs width should be put into the global context of the Higgs precision program. 
Addressing this issue is the primary purpose of this work. We present the results of a systematic study on the Higgs physics potential of a resonant muon collider Higgs factory.

The 125\,GeV muon collider is at an interesting position on the roadmap of future high energy physics experiments. 
A future muon collider can potentially reach center-of-mass energy up to tens of TeV,\footnote{For a review on the physics potential of a high energy muon collider, see, {\it e.g.}, Refs~\cite{Barger:1995hr,Barger:1996jm, Ankenbrandt:1999cta,Delahaye:2019omf,AlAli:2021let}.}, thus providing an unprecedented potential in probing new physics beyond the Standard Model (SM). 
A possible first stage at around 125\,GeV could be strategically advantageous from physics and accelerator perspectives. Such a run provides both a scan of the Higgs resonance, which directly determines the Higgs width and precision Higgs measurements with a production channel ($\mu^+\mu^-\to H$) different from all other collider scenarios. 
As such, it could serve as a Higgs factory at least to a large extent and provide precision determinations on the Higgs couplings and width. 
If one or several $e^+e^-$ Higgs factories are constructed, this will not render the 125\,GeV muon collider useless. On the contrary, as we will show later, there is a strong synergy between a 240\,GeV $e^+e^-$ and a 125\,GeV muon collider due to their different production channels. A combination of the two provides significantly better results than individual ones.    
A crucial question is whether a 125\,GeV first stage is still valuable given the planned high energy runs. 
This is not an easy question to answer and requires both accelerator considerations, technological developments, risk analysis, and detailed analyses of the Higgs measurements in the high energy runs.

Higgs is produced at a 125\,GeV muon collider through an $s$-channel resonance, thus producing a high resonant cross-section of 70~pb. The beam energy spread and initial state radiation reduce the on-resonance production rate to about 22~pb and broaden the peak when performing a lineshape scan~\cite{Greco:2016izi}.~\footnote{These broadening effects can be a double-edged sword. Although their reduces the on-shell production rate and makes it harder to extract the Higgs width information. They allow for a faster prescan~\cite{Conway:2013lca} to locate the Higgs mass pole and also enables resonance production without scanning through the radiative return process~\cite{Chakrabarty:2014pja}.} Such a lineshape scan by changing the muon collider center-of-mass-energy allows for precision extraction of the Higgs width~\cite{Barger:1996jm,Han:2012rb,Conway:2013lca}. However, the instantaneous luminosity expected for a resonant muon collider Higgs factory is only $\mathcal{O}(10^{-2})$ compared to a typical electron-positron collider. One does not expect much about the Higgs precision 
reach at muon colliders compared with other electron-positron Higgs factories. 

We note that although a muon collider Higgs factory would only have $\mathcal{O}(5-20\%)$ statistics compared to electron-positron colliders, the precision on Higgs couplings can be similar due to various reasons. The future electron-positron Higgs factories generally achieve $\mathcal{O}(1\%)$ precision on Higgs couplings, despite the number of Higgs produced would be around one million~\cite{Baer:2013cma,An:2018dwb,Abada:2019zxq}.\footnote{One exception is the $HZZ$ coupling which is determined from an inclusive measurement of the Higgs cross-section through the recoil mass technique.} This mismatch (except for $HZZ$ coupling) between statistics and the achievable precision is driven by the lack of precision on the Higgs total width. 
One can understand it as all exclusive cross-sections are measured and parameterized as the rescaling of coupling produced divided by the total width for on-shell Higgs bosons. The uncertainty of the Higgs width propagates to the determination of couplings, regardless of whether it is a free parameter, in a ``model-independent fit", or a derived quantity, in a constrained fit or an EFT fit. For instance, at CEPC~\cite{An:2018dwb,CEPCStudyGroup:2018ghi}, the Higgs width precision is around $2.8\%$ when the width is a free parameter and only improves to be around $2.4\%$ when the width is not a free-parameter but as an internal parameter through the error propagation of the different partial widths. 

While the current (Snowmass muon collider forum) benchmark integrated luminosity for the 125\,GeV muon collider is 20~$\abi$~\cite{lumibenchmarkSnowmass}, we also provide projections for 5~$\abi$ as an alternative scenario and conservative estimate.
We show that a future 125\,GeV muon collider would provide around $6.8\times 10^{4}$ to $2.7\times 10^{5}$ thousand of Higgs bosons, with a $\mathcal{O} (\%)$ Higgs width precision from a lineshape scan. These many Higgs bosons together with such a measurement for the total width would allow the muon collider to achieve a Higgs coupling precision at the percent level.

The rest of our study is organized as follows. In \autoref{sec:width} we present the systematic study on the Higgs width determination from lineshape scan. Our study here includes beam energy spread and initial state radiation and contains various further considerations. We clarify the fit inputs, fit procedure, and develop an optimized scan strategy. In \autoref{sec:meaasurements} we study the exclusive channels and their precision. These results enable us to perform global fits that provide the holistic picture of the Higgs program in various scenarios in \autoref{sec:couplings}. Finally, we summarize in \autoref{sec:conclusion}.

%=========================

\section{Width Determination}
\label{sec:width}

The Higgs width at a 125~GeV muon collider is uniquely determined through a lineshape mapping process. 
The potential sensitivity to the Higgs width has been explored in previous studies. However, so far, a complete treatment of various effects is still missing, hindering the possible extraction of Higgs program precision in a muon collider resonant Higgs factory. In this section, we carry out the width determination comprehensively.

\subsection{Higgs Lineshape at Resonant Muon Collider Higgs Factory}
\label{sec:widthlinshape}

The observed cross-section at a given beam center-of-mass energy $\overline E_{\rm com}$ is the convolution of three effects, the beam energy spread that depends on the beam quality, the initial state radiation from QED, and the kernel of a Breit-Wigner distribution that depends on the Higgs mass, width and coupling strength. In general, the measured cross-section at a beam center-of-mass energy $\overline E_{\rm com}$ can be expressed as
\bea
    \sigma_{\mu^+\mu^- \to XX}(\overline E_{\rm com}, \sigma_E, m_H, \Gamma_H, \hat\upmu)=&\label{eq:convolutionlineshape}\\
    \int d E_{\rm com} d x~F_{\rm beam}(E_{\rm com};& \overline E_{\rm com}, \sigma_E)\times F_{\mu^+\mu^-}^{\rm ISR}(x; E_{\rm com}^2)\times\hat\sigma_{\mu^+\mu^- \to XX}(x^2 E_{\rm com}^2; m_H, \Gamma_H, \hat \upmu),\nonumber
\eea
where the different distributions and parameters are described in what follows.

The core distribution depending on the Higgs boson physical properties is the Breit-Wigner distribution,
\bea
\sigma_{\mu^+\mu^-\to XX}(\hat s; m_H, \Gamma_H, \hat \upmu)&=&\frac {4\pi \hat\upmu \Gamma_H^2{\rm BR^{SM}}(H\rightarrow \mu^+\mu^-) {\rm BR^{SM}}(H\rightarrow XX)} {(\hat s-m_H^2)^2+\Gamma_H^2m_H^2} +\sigma^{\rm SM}_{\rm bkg}{(\mu^+\mu^-\to XX)}.
\label{eq:partonlevelxs}
\eea
Here $\sigma^{\rm SM}_{\rm bkg}{(\mu^+\mu^-\to XX)}$ represents all the SM background processes that do not involve the Higgs amplitudes. The SM background processes include irreducible and reducible backgrounds that fake the $XX$ final state. In this treatment, the rate-changing interference effects proportional to the real part of the propagator can be safely ignored. Such interference effect is suppressed by $\Gamma_H/m_H$ and the helicity due to the mismatch between SM background process and Higgs process. The on-shell rate-changing effects proportional to the absorptive part of the propagator are also negligibly small, as all the leading processes here do not invoke a strong or weak phases for SM Higgs~\cite{Campbell:2017rke}. $\Gamma_H$ is the Higgs total width, and ${\rm BR^{SM}}$ denotes the SM Higgs decay branching fractions to a given state. $m_H$ is the Higgs pole mass, and $\hat s$ is the actual collision center-of-mass-energy squared, considering the beam energy spread and initial state radiation effects. In generating the (pseudo) experimental data, we assume all parameters are at their SM value, and hence $m_H=125$~GeV, $\Gamma_H=4.1$~MeV, and the signal strength scaling factor $\hat \upmu=1$. However, in the fitting procedure, all these three parameters are set as free parameters to determine the power of the lineshape scan. We describe the fitting procedure in detail in the next subsection. 

The beam energy profile function captures beam energy spread (BES) from beam dynamics, which is assumed to be Gaussian,
\beq
F_{\rm beam}(E_{\rm com}; \overline E_{\rm com}, \sigma_E)=\frac 1 {\sqrt{2\pi}\sigma_E}\exp\left[-\frac {(E_{\rm com}-\overline E_{\rm com})^2} {2\sigma_E^2}\right],
\eeq
where $E_{\rm com}$ is the beam-delivered collision center-of-mass energy that the beam provides, and $\overline E_{\rm com}$ is what we typically refer to as collider energy. For instance, a 125~GeV muon collider implies our target mean energy of the collision, $\overline E_{\rm com}=125$~GeV. Furthermore, $\sigma_E$ represents the standard deviation of the beam energy profile. A precise determination of BES is critical to successfully extract the Higgs boson width, as the Higgs width precision will also be subject to uncertainties in $\sigma_E$. Hence, BES will contain a systematic uncertainty that directly propagates to the final results. One can only evaluate it with the help of beam physicists, which will be the subject of a future study.

The Initial State Radiation (ISR) effects, especially with higher-order effects of multiple soft-photon emission, are taken into account beyond two-loop effects, using the structure-function. Such treatment is discussed in detail in Ref.~\cite{Greco:2016izi}, and various formalisms and approximations are compared. In this work, without loss of accuracy, we adopt the Jadach-Ward-Was~\cite{Jadach:2000ir} formalism (b)~\footnote{This is consistently checked and recommended in Ref.~\cite{Greco:2016izi}.}, where
\beq
F_{\mu^+\mu^-}^{\rm ISR}(x;\hat s)=\exp\left[\frac {\beta_\mu} {4}+\frac {\alpha}{\pi} \left(-\frac 1 2+\frac {\pi^2}{3}\right)\right]\frac {\exp[-\gamma\beta_\mu]} {\Gamma[1+\beta_\mu]}\beta_\mu (1-x)^{\beta_\mu-1}\left(1+\frac{\beta_\mu}{2}-\frac 1 2 (1-x)^2\right),
\eeq
with $\beta_\mu$ the common loop factor
\beq
\beta_\mu=\frac {\alpha}{\pi} \left(\log\frac {\hat s}{m_\mu^2} -1 \right).
\eeq
In the above equations, $\gamma$ is the Euler-Mascheroni constant and $\Gamma[x]$ is the Gamma function, while $\alpha$ is the fine-structure constant. Physically, $x$ represents the fractional energy from a given beam-delivered collision center-of-mass energy $E_{\rm com}$ that can be utilized for the hard collision process of $\mu^+\mu^-\rightarrow H\rightarrow XX$.

\begin{figure}
\begin{center}
\includegraphics[width=0.49\linewidth]{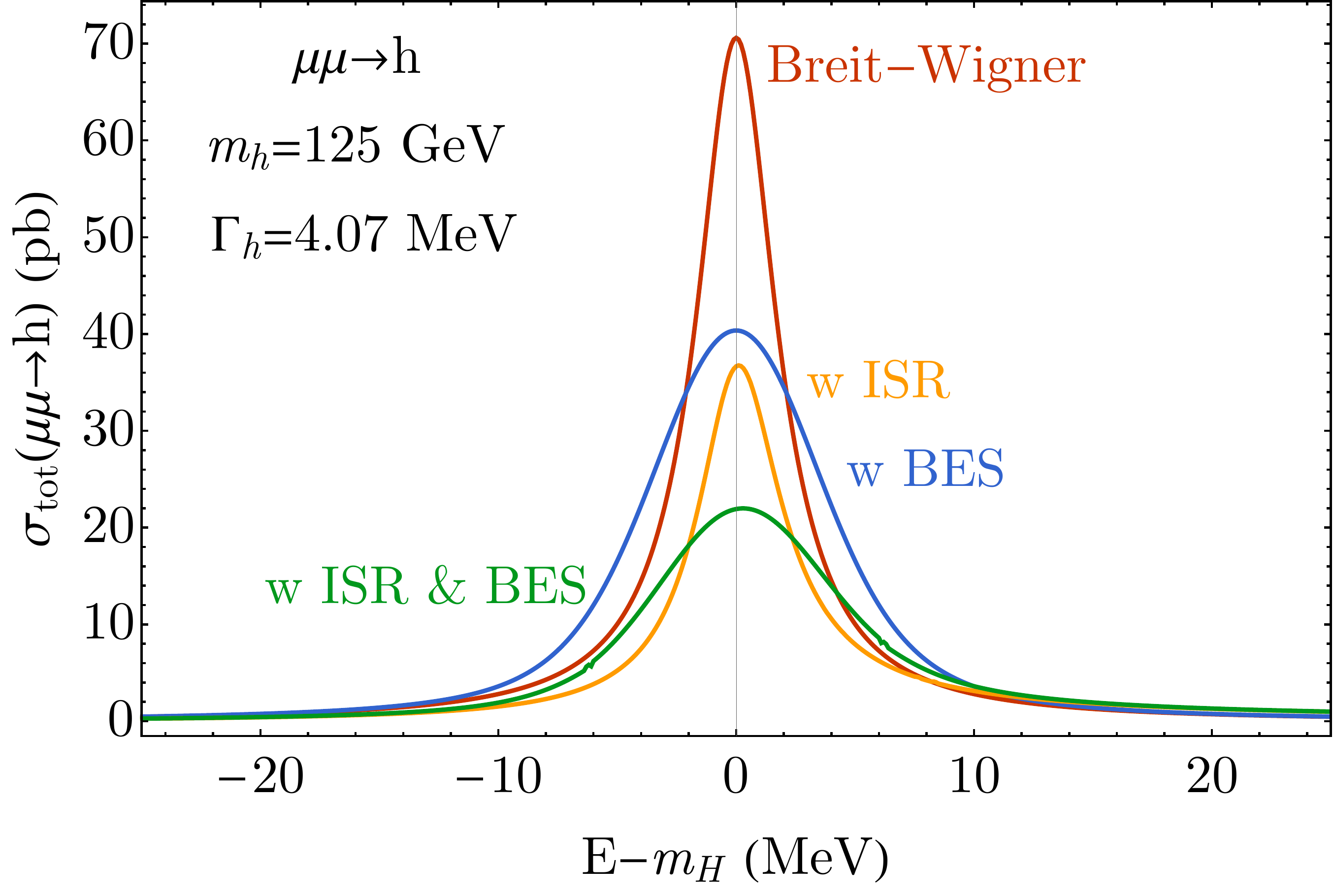}
\includegraphics[width=0.49\linewidth]{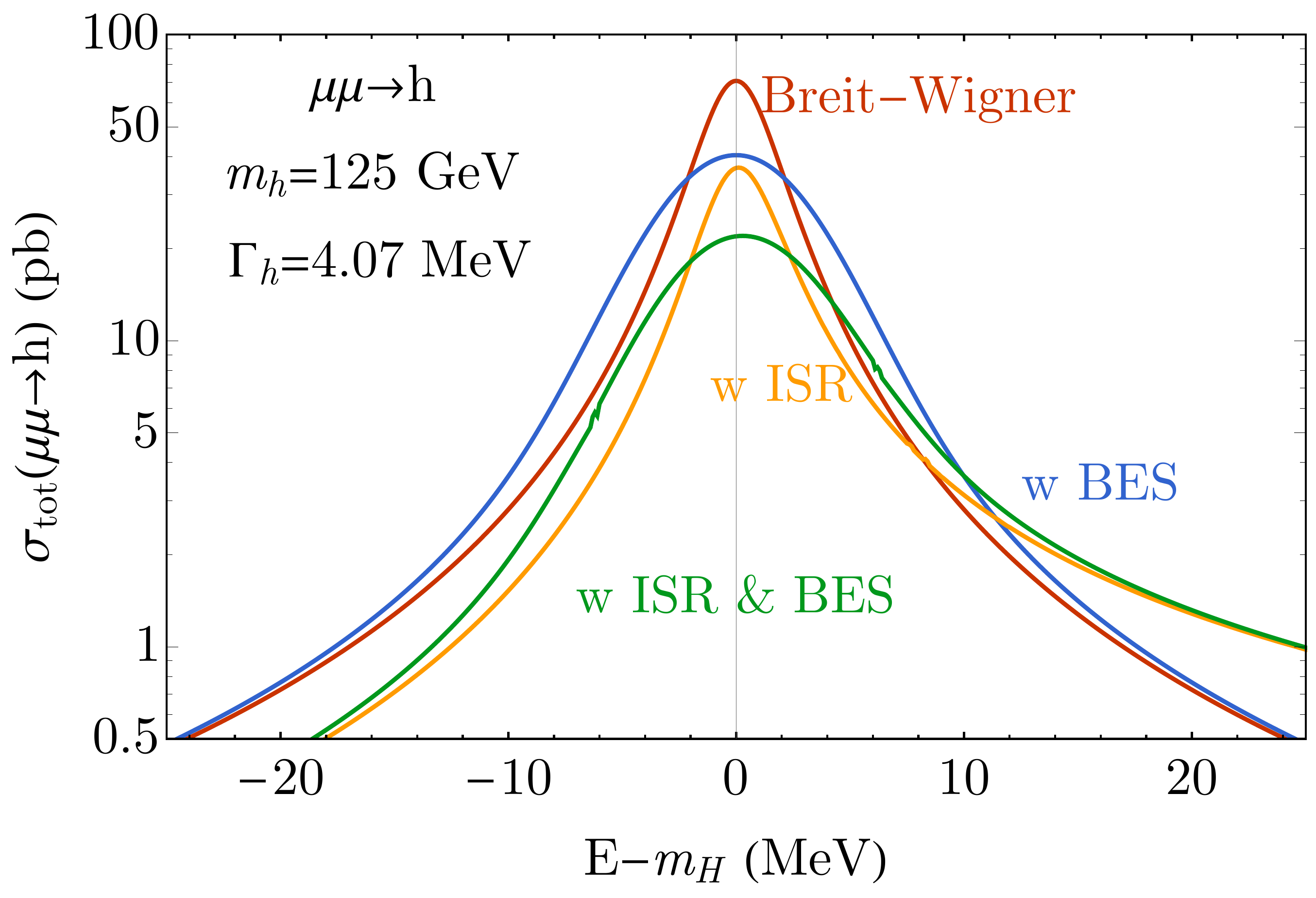}
\caption{The Higgs lineshapes with various effects as a function of $E_{\rm com}-m_H$, the actual hard scattering center-of-mass energy difference with respect to the Higgs pole mass. Here $E$ is $x E_{\rm com}$, $\delta(x-1)x\overline E_{\rm com}$, $x \overline E_{\rm com}$, and $\overline E_{\rm com}$ for the Breit-Wigner, Breit-Wigner plus BES, Breit-Wigner plus ISR, and Breit-Wigner with both ISR and BES, respectively.\footnote{Note here the less dependence on convolution parameters, $x$, $E_{\rm com}$, the closer to the final, post-convolution distribution.} The detailed meanings of these quantities are described \autoref{sec:widthlinshape}. We show the theoretical predictions in linear ({\em left panel}) and logarithmic ({\em right panel}) scales. To obtain physical observable, one needs to decide on the scan range, separation, and luminosity assignment of each scan step. These will form a discrete set of event counts that are further subject to statistical fluctuations, forming the pseudo experimental data set to feed into the Higgs fit.}
\label{fig:lineshape}
\end{center}
\end{figure}

Such a complex behavior forbids us from complete analytic control, and we then turn to numerical simulations. However, a few critical semi-analytic understanding impacts our modeling of the Higgs precision physics in various parts throughout this paper. We show the convoluted distributions of each effect in \autoref{fig:lineshape} and comment on their features here.
The ISR effects {\it does not} broaden much the Higgs lineshape near the peak region. Instead, it reduces the Higgs peak height, creates an asymmetry in the lineshape, and redistributes the on-peak cross-section to higher $E_{\rm com}$ regimes. This effect can be seen comparing the yellow lines with the red lines \autoref{fig:lineshape}, as the ISR effect is a sharp one-sided distribution with respect to $x=1$. 
The beam energy spread effects broaden the Higgs lineshape near the peak region, as seen by comparing the blue curves with the red curves in \autoref{fig:lineshape}. Very close to the peak region, the Gaussian beam energy spread and the Breit-Wigner distribution differ at higher order. The Full-Width-Half-Maximum (FWHM)\footnote{We can use FWHM as an approximate measure of the width of the distributions.} for a Gaussian beam energy spread of $\sigma_E$ is $2\sqrt{2\log 2}\sigma_E\simeq 2.3 \sigma_E$, while the FWHM for the Breit-Wigner is $\Gamma_H$. Hence the broadening is dominated by the bean energy spread for the current benchmark of the muon collider beam property. On the other hand, if there are systematic uncertainties of $\sigma_E$ comparable to that of the determined Higgs width, we will need to consider more information in the regions slightly further away from Higgs mass pole. In this sense, our scanning region of $\pm 8~\mev~\simeq 4\Gamma_H$ ensures robustness to extract the Lorentzian (Breit-Wigner) width against Gaussian smearing. 

\subsection{Lineshape Observables}

Having understood Higgs lineshape behavior at the muon resonant Higgs factories, we can proceed and discuss the observables concerning Higgs width determination and the corresponding fitting procedure. The scanned lineshape, depending on how finely one changes the beam energy $\overline E_{\rm com}$, is a collection of measurements
\beq
\{\sigma(\overline E_{\rm com}^i, \sigma_E, m_H, \Gamma_H, \hat \upmu)\},~{\rm for\ a\ set~of\ }\{\overline E^i_{\rm com}\}
\eeq
and for a given assignment of the luminosity, $\{L^i\}$, to be collected at each energy point. With these, one obtains the uncertainty $\Delta\sigma$ for each measurement point to map out the Higgs lineshape
\beq
\{\sigma(\overline E_{\rm com}^i, \sigma_E, m_H, \Gamma_H, \hat \upmu), \Delta\sigma(\overline E_{\rm com}^i, \sigma_E, m_H, \Gamma_H, \hat \upmu)\},~{\rm for\ a\ set~of\ }\{\overline E^i_{\rm com}, L^i\}
\eeq
for each process of $\mu^+\mu^-\to H\to XX$.

Given that no (non-Higgs) SM background rate will vary much\footnote{Non-resonant SM rate will not vary more than $\mathcal{O}(10~\mev/125~\gev)$ within the scan range of $\pm 8~\mev$.}, one can collectively check the total number of signal Higgs produced for a given scanning strategy, while the background being independent of it. For a reasonable scanning of equal luminosity, the effective Higgs cross-section $\sigma_{\rm eff}$ is then the average of $\{\sigma(\overline E_{\rm com}^i, \sigma_E, m_H, \Gamma_H, \hat \upmu)\}$, being
\beq
\sigma^{\rm eff}_{\mu^+\mu^-\rightarrow XX} \equiv \frac {\sum_i \sigma_{\mu^+\mu^-\rightarrow XX}(\overline E_{\rm com}^i, \sigma_E, m_H, \Gamma_H, \hat \upmu)\times L^i} {\sum_i L^i}.
\label{eq:xseffectivedefinition}
\eeq
This $\sigma^{\rm eff}_{\mu^+\mu^-\rightarrow XX}$ provides the equivalent cross-section for a given process of $\mu^+\mu^-\rightarrow XX$, a useful quantity for the coupling precision fits. For the SM Higgs, with a beam spread $0.003\%$ and 125 GeV center-of-mass energy we can find $\sigma_E=2.7$~MeV.
\footnote{Each beam of 62.5 GeV energy vary by 0.003\%, independently.}
As argued earlier, to effectively extract the Higgs width, as well as avoiding losing too much valuable luminosity in in the regime with little Higgs signal, we need to cover an energy regime of the order $\mathcal{O}(2.3\sigma_E\oplus \Gamma_H)=\mathcal{O}(7~\mev)$.

The discussion above and the previous subsection also lead to another important approximation we can make for individual exclusive Higgs rate measurements, which is the subject of the next section and is the critical input for the Higgs coupling fit. For fine-enough scanning strategy\footnote{We will discuss the impact of different scanning strategies in the following subsection numerically.}, the effective Higgs cross-section is closer to a narrow-width-approximated rate, than an exact on-shell rate. The critical difference between these two choices of the effective cross-section is their width dependence, the former $1/\Gamma_H$ and the latter $1/\Gamma_H^2$. The key argument is that the scanning range and the effective width of the BER with ISR effect are both a factor of a few larger than the Higgs width. Hence, we effectively integrated over the lineshape that gets rid of one power of $\Gamma_H$, in which it resembles the zero-width approximation. 
Then, for our scanning strategy, we can parameterize the exclusive Higgs rate (following \autoref{eq:effectivexs}, and after subtracting SM background in \autoref{eq:convolutionlineshape} and \autoref{eq:partonlevelxs}) to be 
\beq
\sigma^{\rm eff}_{\mu^+\mu^-\to H\to XX}= \eta^{\rm eff} \left(\frac {\Gamma_H^{\rm SM}} {\Gamma_H}\right) \sigma_{\mu^+\mu^- \to H \to XX}(m_H, \sigma_E, m_H, \Gamma^{\rm SM}_H, 1).
\label{eq:onshellxs}
\eeq
Here $\sigma_{\mu^+\mu^- \to H \to XX}(m_H, \sigma_E, m_H, \Gamma_H, 1)$ represents when the beam energy is on the Higgs pole $\overline E_{\rm com}=m_H$ for the Higgs process, as defined in \autoref{eq:convolutionlineshape}, after removing the flat SM background. For channels we do not rely on to extract the Higgs width from lineshape fitting, the total number of collected signal Higgs events will approximately be inversely proportional to Higgs width. A constant effective parameter $\eta^{\rm eff}$ describes such dependence that relies on the scanning strategy $\{\overline E_{\rm com}^i, L^i\}$. For the scanning range of $m^{\rm SM}_H\pm 8~\mev$ with 11 steps of even integrated luminosity that is adopted eventually in this study, the effective parameter 
\beq
\eta^{\rm eff}=0.615. 
\label{eq:effectivexs}
\eeq
Such value of $\eta^{\rm eff}$ implies that scanning, instead of directly sitting on the resonance (where $\overline E_{\rm com}=m_H$, and convoluted with ISR and BES), reduces the total number of Higgses by about 38\% to gain the knowledge of Higgs width. This treatment simplifies the Higgs coupling and EFT fitting significantly while not losing the major correlations in the lineshape fitting. 
We note that this approximation is clearly not the zero-width approximation, given that our scanning measurement resolves the lineshape and is {\it differentially} sensitive to the Higgs width.

\subsection{Lineshape Scan Considerations}
\label{sec:scanconsiderations}

The {\it leading channels} for precision measurements of the Higgs at a 125 GeV muon collider would be $\mu^+\mu^-\to H\to b\bar b$ and $\mu^+\mu^- \to H\to WW^*$, from the large signal statistics and relatively low SM background. For the Higgs lineshape fit, one can focus on these two channels. A detailed analysis of these channels for signal and background can be found in the next section. Importantly, these two channels also example how the fitting results and scaling vary for different signal background ratios. For $\mu^+\mu^-\to H \to b\bar b$, the signal background ratio is around 2:5. For $\mu^+\mu^- \to H \to WW^*$, the signal background ratio is around 50:1, depending on the final states of the $WW^*$ channels. We can then generate pseudo experimental data, perform the fitting, and optimize the process.

Before designing the scanning range, we note that we do not know the Higgs mass {\it a priori} to the level of a few MeV. A {\it prescan} is needed. We expect to know the Higgs mass to a precision of $\mathcal{O}(10~\mev)$ at the LHC and future lepton colliders. 
Hence, at a 125 GeV muon collider, we would need to spend some luminosity to perform a prescan to determine the Higgs pole location, ideally to $\mathcal{O}(\mev)$. Fortunately, thanks to the high resonant cross-section, a few hundred pb$^{-1}$ of integrated luminosity for such a prescan would be sufficient~\cite{Conway:2013lca}. In this work, we focus on discussing the Higgs property determination post such prescan. Further optimization of the luminosity spending plan to be explored in future works.  

The {\it scan range and luminosity per scanning step} directly impacts the outcome of the lineshape fit. 
With the three free parameters given by the Higgs mass, $m_H$, total width, $\Gamma_H$, and signal rate $\hat\upmu$, as indicated in \autoref{eq:convolutionlineshape}, the optimal scan strategy requires a balanced on-peak and off-peak luminosity spending budget. As we shall see later, among these three parameters, $\Gamma_H$ and $\hat \upmu$ are strongly correlated but independent from $m_H$. Near the peak region, the result is anticipated to compensate the rate loss from larger $\Gamma_H$ with larger $\hat \upmu$. One shall note that, if one fixes $\hat\upmu$, the width information is then optimally captured at $\overline E_{\rm com}=m_H$~\footnote{Strictly speaking, it will require $\overline E_{\rm com}$ to be marginally above $m_H$ due to the ISR effect.}, using Fisher information. We argue that a consistent extraction of the Higgs properties should require both $\Gamma_H$ and $\hat \upmu$ (as well as $m_H$) as free parameters. This requirement is because we will not know these parameters to a precision much better than from a resonant muon collider Higgs factory from any other collider that have been envisioned so far in reasonable fitting frameworks. 

While one can calculate the Fisher information and determine the optimal scanning strategy $\{\overline E^i_{\rm com}, L^i\}$, we chose {\it not to} to avoid unphysical optimizations for effects that we have not yet taken into account for this work. These effects are mainly, 1) the unknown Higgs mass $m_H$ at sub-MeV level (even with a prescan), 2) the un-modeled/unknown effects of the beam energy profile uncertainties. We want to have a safe margin in the scanning strategy, and hence assume in our study an equal distance in scanning collider energy $\overline E_{\rm com}^i$ and equal luminosity per step $L^i$. Yet, the proper scanning range and number of scanning steps are unknown, and we provide an answer in this study.

\subsection{Lineshape Fit and Results}
\label{sec:widthresults}

We make the standard assumption that future experimental data matches the SM predictions. For each choice of scanning range, and luminosity per step, $\{\overline E^i_{\rm com}, L^i\}$, we generate the pseudo experimental data following Poisson distributions whose central values are calculated according to SM predictions following \autoref{eq:convolutionlineshape}. These central values also fold in signal and background selection and (including geometrical) acceptance efficiencies $\epsilon_{s}$ and $\epsilon_{b}$, respectively, that are obtained in the next section. These values corresponds to 
\bea
\epsilon^{b\bar b}_{b}\sigma^{\rm SM}_{\rm bkg}(\mu^+\mu^-\to b\bar b)&=& 6.2 {\rm~pb},\nonumber\\ \epsilon^{b\bar b}_{s}\left(\sigma_{\mu^+\mu^- \to b\bar b}(m^{\rm SM}_H, 2.7~\mev, m^{\rm SM}_H, \Gamma^{\rm SM}_H, 1)-\sigma^{\rm SM}_{\rm bkg}(\mu^+\mu^-\to b\bar b)\right)&=& 14.5 {\rm~pb},\\
\epsilon^{WW}_{b}\sigma^{\rm SM}_{\rm bkg}(\mu^+\mu^-\to WW^*)&=& 2.1 {\rm~pb},\nonumber\\ \epsilon^{WW}_{s}\left(\sigma_{\mu^+\mu^- \to WW^*}(m^{\rm SM}_H, 2.7~\mev, m^{\rm SM}_H, \Gamma^{\rm SM}_H, 1)-\sigma^{\rm SM}_{\rm bkg}(\mu^+\mu^-\to WW^*)\right)&=& 47{\rm~fb}.
\eea

To signify the correlation between Higgs total width $\Gamma_H$ and the flat direction of the exclusive rate at all the other colliders\footnote{In other words, the hunt for Higgs width precision is to fight against the uncertainties associated with this general direction. At the LHC, we can try to use the on-shell off-shell rate ratio (with an assumption that coupling do not change over different scales)~\cite{Kauer:2012hd,Caola:2013yja}, (nearly) on-shell diphoton mass shift~\cite{Dixon:2013haa} and on-shell diphoton rate change~\cite{Campbell:2017rke} to help constrain width. At future $e^+e^-$ machines, we can use the inclusive $ZH$ associated production, together with other exclusive channels to constrain width.}, instead of fitting the product of coupling strength $\hat\upmu$, we fit a rescaled strength along this flat direction, 
\beq
\tilde \upmu \equiv \hat \upmu \left(\frac {\Gamma_H^{\rm SM}}{\Gamma_H}\right).
\eeq

To explore various properties of the fit, we generate around 10 million sets of pseudo experimental data according to the above procedure for various scanning strategies $\{\overline E^i_{\rm com}, L^i\}$. Then, we generate dense lists of lineshape data following \autoref{eq:convolutionlineshape} for a range of Higgs total width between $0.1\%$ and $300\%$ of $\Gamma_H^{\rm}$ and shift them accordingly to $m_H$ and $\tilde \upmu$. 
This allows us to build a likelihood function responding to each pseudo experimental data. For each set of pseudo experimental data, we find the best fit parameter for   $\{\Gamma^H, m_H, \tilde \upmu\}$. We obtain a set of the best-fit parameters for the pseudo experimental data and hence can study uncertainties associated with them. Through marginalization, we can study the individual parameter's uncertainties for a given scanning strategy $\{\overline E^i_{\rm com}, L^i\}$, as well as their correlations. We report our findings next. Since the width is the parameter of major interest here, we use the width uncertainty to show various properties in the following.

\begin{figure}
\begin{center}
\includegraphics[width=0.7\linewidth]{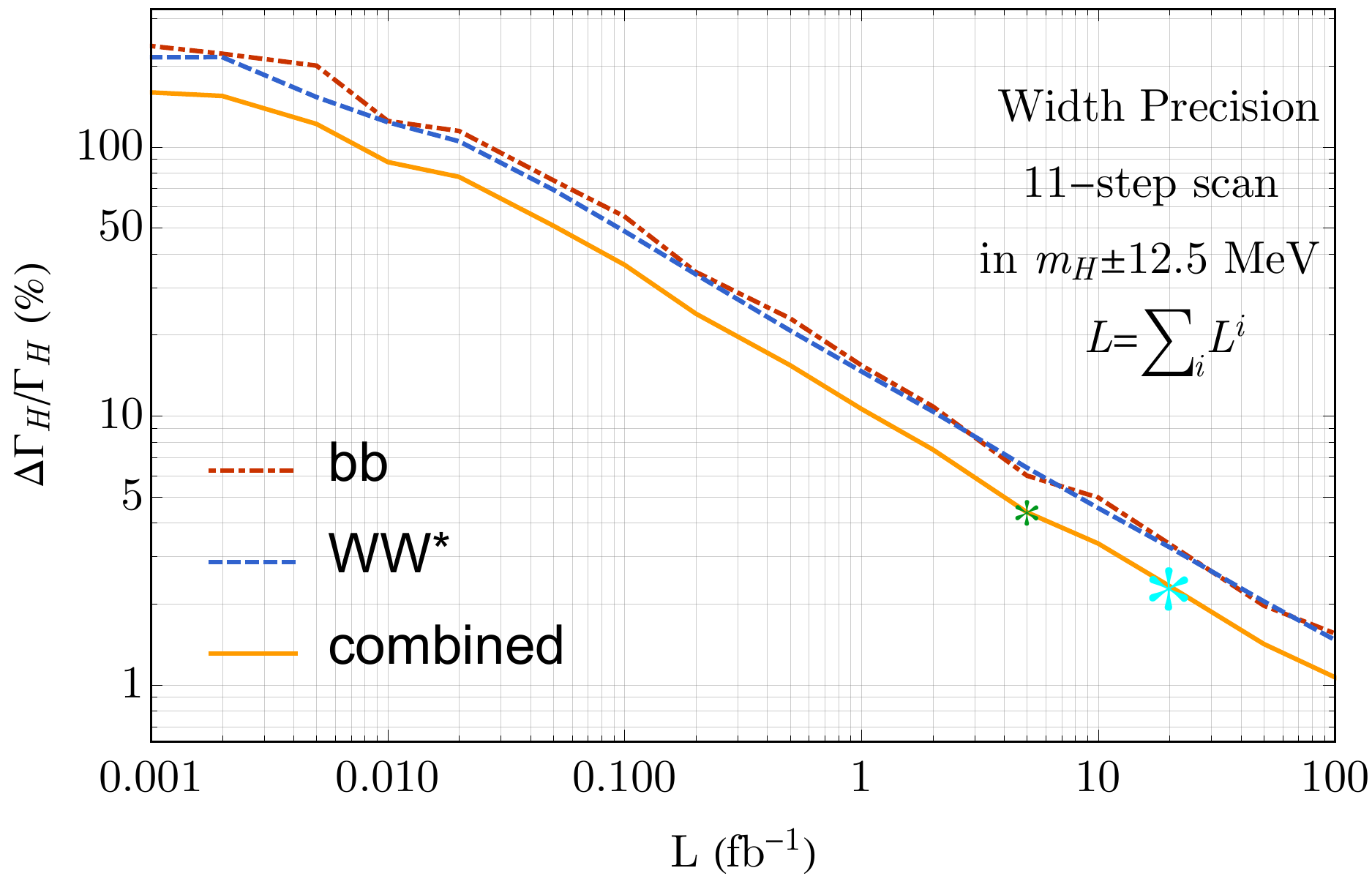}
\caption{The projected width sensitivity as a function of total luminosity for the lineshape scan. We also highlight two benchmarks of integrated luminosity considered in this study in green and cyan asterisk symbol.\footnote{Note that this scanning range of $m_H\pm 12.5~\mev$ is sub-optimal compared to our final scanning range of $m_H\pm 8~\mev$. However, this does not affect our discussion here, which is the precision scaling with luminosity. The precision dependence on the scan range is discussed next.}
\label{fig:widthlumi}}
\end{center}
\end{figure}

Understanding {\it precision scaling with luminosity} helps with our planning for various scenarios. We avoid assuming linear expansion in determining the central values and uncertainties associated with the fit from the descriptions above. Hence, it would be helpful to check how the resulting precision on the fitted parameter, e.g., the normalized standard deviation of width $\Delta \Gamma_H/\Gamma_H$, scales with the integrated luminosity at the 125 GeV muon collider. On the other hand, since the lineshape function is a regular function of $\Gamma_H$, we do expect the precision of the determination of the width to scale as $1/\sqrt{L}$ where $L$ is the integrated luminosity. In \autoref{fig:widthlumi} we show how our fit result depends on total integrated luminosity for an 11-steps scan within $m_H^{\rm SM}\pm 12.5~\mev$ (the dependence on the scanning range and the number of steps of scanning are discussed next). We can see from the figure that the width precision does scale as $1/\sqrt{L}$, which is a useful verification and input for planning practices where the achievable integrated luminosity for a 125 GeV muon collider might develop and vary. In the extremely low luminosity part, the scaling fails for two reasons: the fitting is no longer linear, and the theory lineshape space we generated are truncated at large Higgs widths. It is also interesting to note that, although signal statistics are quite different between these two different leading channels, they contribute approximately equally to the width determination. In other words, although $WW^*$ signal is a factor of $\sim 4$ small in signal events (post selection efficiencies), due to the negligible background, has a similar statistical power in width determination compared to $b\bar b$. 

\begin{figure}
\begin{center}
\includegraphics[width=0.7\linewidth]{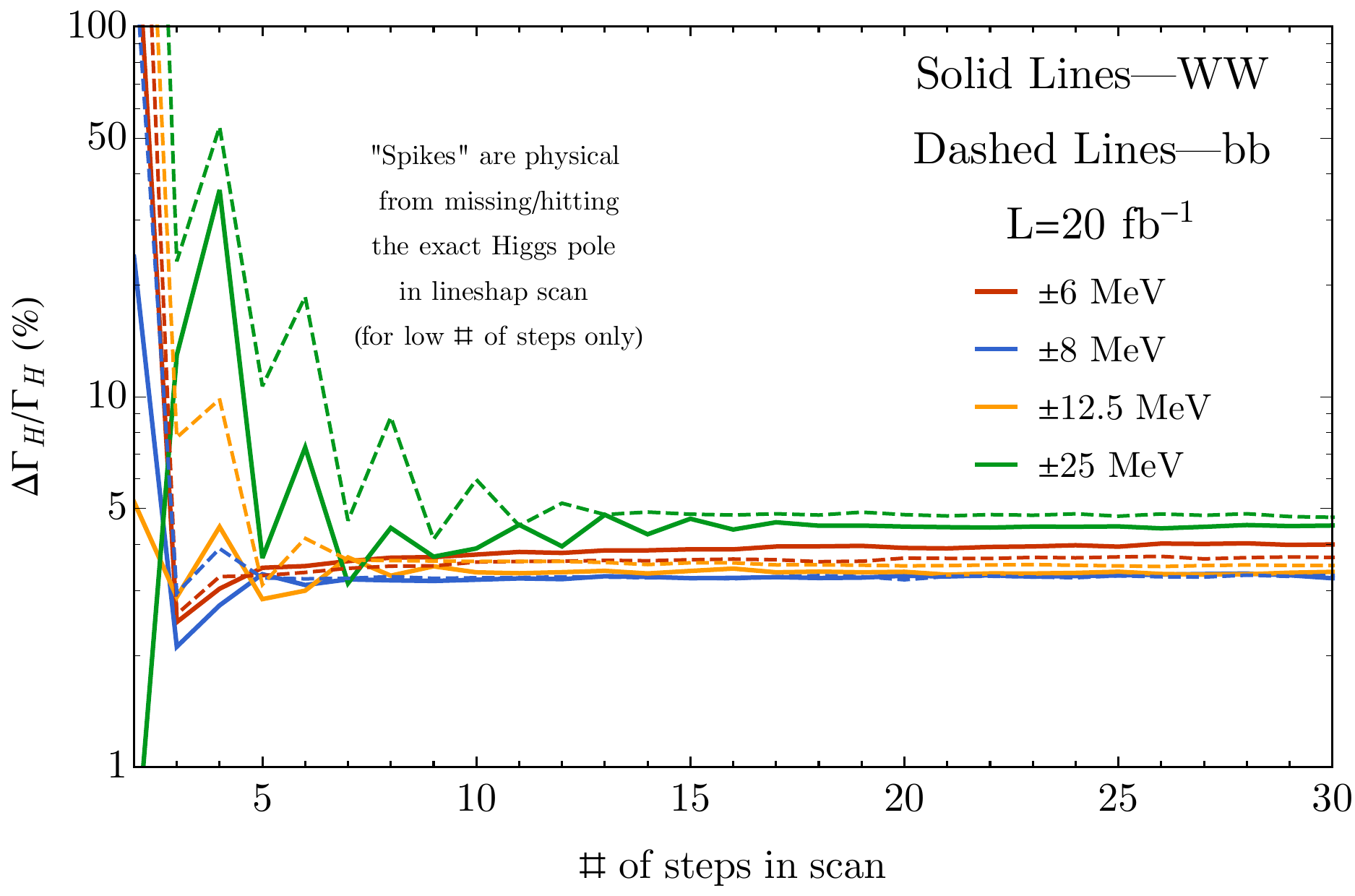}
\caption{The projected width sensitivity as a function of scanning steps for various choices of scanning ranges around the Higgs pole mass $m_H$, with a fixed total luminosity of $20~\fbi$.
\label{fig:widthsteps}}
\end{center}
\end{figure}

{\it Precision depends on the scan range and number of scan steps.} For a fixed total integrated luminosity, the scan range, step size, and luminosity assignment determine the outcome of the lineshape scan. We show the dependence on the number of scan steps for various scan ranges in \autoref{fig:widthsteps}. If the scan range is too broad, one effectively wastes the majority of integrated luminosity without the Higgs signal. In this case, the obtained data contains no information about the Higgs width, apart from validating the calculable SM background. Even worse, in the large scan range scenarios, if the scanning steps are low, one may not have sampled enough data in the Higgs peak region, resulting in considerable uncertainty in the determination of the Higgs parameters. One can see in \autoref{fig:widthsteps} the green curves have worse precision compared to other scan ranges, and as well ``spikes'' for missing or hitting the Higgs pole. For a high enough number of scanning steps, one does not have ``spikes'' in any scan ranges. For scan range $\pm 12.5~\mev$, $\pm 8~\mev$, and $\pm 6~\mev$, ten steps are sufficient to avoid such fluctuations in precision, as one sample enough points around the peak to map out the Higgs width information. However, when the scan range is too narrow, as argued in the previous subsections, one cannot separate the overall reduction of the on-shell rate due to a broader $\Gamma_H$ from a change in signal rate $\tilde \upmu$. 

Note that focusing on too narrow a region around the Higgs pole, one would lose the ability to separate effects such as BES and ISR, making a systematic cross-check hard. Such effect is reflected in comparing the blue and the red curves. We can see that, interestingly, one gets consistently worse precision on the Higgs width with a narrower range of $\pm 6~\mev$ (red curves), although one effectively increased the total number of Higgs bosons produced due to the closer focus on the pole region. 

This understanding of the lineshape fit allows us to propose a (reasonably\footnote{As discussed in various places in this section (e.g., \autoref{sec:scanconsiderations}), here we want to avoid over-optimization and save room for additional systematics}) optimized scan strategy here, we take
\beq
{\rm scan~range:}~m_H\pm 8~\mev,~{\rm with~11~steps~of~scan~with~even~spacing~in~energy.}
\label{eq:strategy}
\eeq
Note that the $11$ step is a somewhat random choice. So long as it is larger than 7, we avoid the fluctuations caused by how well one hits the Higgs pole, as shown in blue curves in \autoref{fig:widthsteps}.

\begin{table}[t]%[h]
\centering
\begin{tabular}{|c||c|c|c|c|c|c|} \hline
 &  \multicolumn{6}{|c|}{Precision (w BES and ISR)} \\  \hline
 channel &   $\delta\Gamma_H/\Gamma_H$  &  $\delta\tilde \upmu$ & $\delta m_H$ (MeV) &  \multicolumn{3}{|c|}{correlation $\rho$} \\ \hline\hline
$b\bar b$ & 3.22\% & 1.03\% & 0.31 & 0.762 ($\delta \Gamma_H$-$\delta\tilde \upmu$) & -0.040 ($\delta \Gamma_H$-$\delta m_H$) & -0.037 ($\delta m_H$-$\delta\tilde \upmu$) \\  \hline
$WW^*$ & 2.80\% & 0.84\% & 0.29 & 0.625 ($\delta \Gamma_H$-$\delta\tilde \upmu$) & -0.077 ($\delta \Gamma_H$-$\delta m_H$) & -0.081 ($\delta m_H$-$\delta\tilde \upmu$)\\  \hline
{\bf Combined} & 2.1\% & 0.65\% & 0.21 & \multicolumn{3}{|c|}{} \\ \hline
\end{tabular}
\caption{The width fit results for the two leading channels with the scanning strategy adopted in this study (\autoref{eq:strategy}) for an integrated luminosity of $20~\fbi$. Here we also show the direct combination of the precision from these two channels. Note that when we perform the global fit, these two channels are treated independently and their correlations with couplings will be constrained by other inputs, resulting in a better combined Higgs total width precision.}
\label{tab:widthresult}
\end{table}

With the above choice of scan strategy, we can calculate the expected Higgs precision from the lineshape fit. We report then in \autoref{tab:widthresult} for the two leading channels and report the combined precision and correlations. We note that, when combined with the Higgs exclusive measurements reported in the next section, we carefully avoid {\it double-counting} the $WW^*$ and $b\bar b$ channel. In the final $\kappa$ and EFT fits, the correlated channels from the width determination are treated together as inputs. 
We can see both channels can determine Higgs properties to a great precision from a lineshape scan, each bringing a 3\% level total width precision, 1\% level signal strength, and 0.3~MeV level of Higgs mass precision, for 20 fb$^{-1}$ of total luminosity. 

\begin{figure}
\begin{center}
\includegraphics[width=0.45\linewidth]{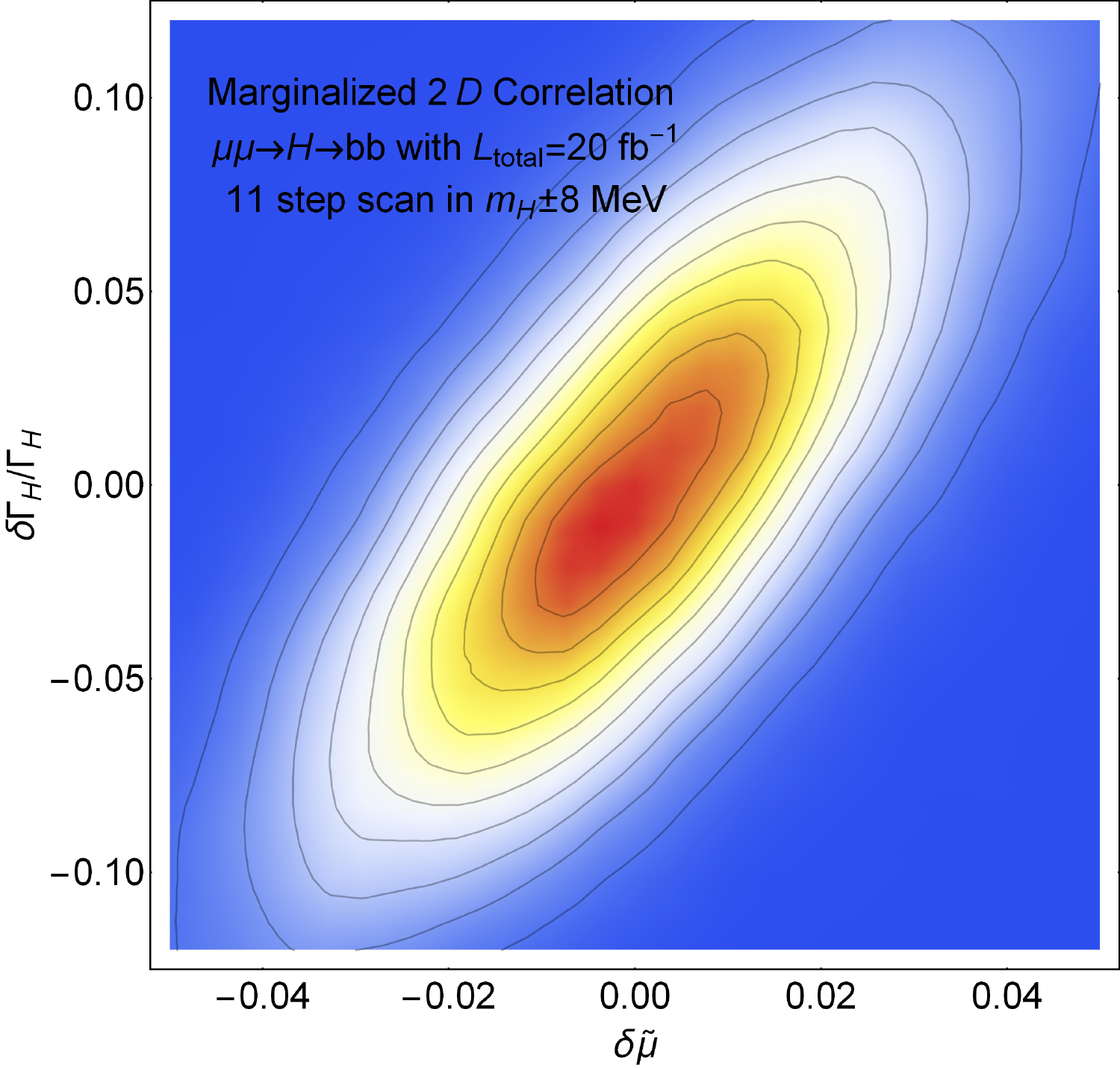}%{Rate_Width_correlation_b\bar b.pdf}
\includegraphics[width=0.45\linewidth]{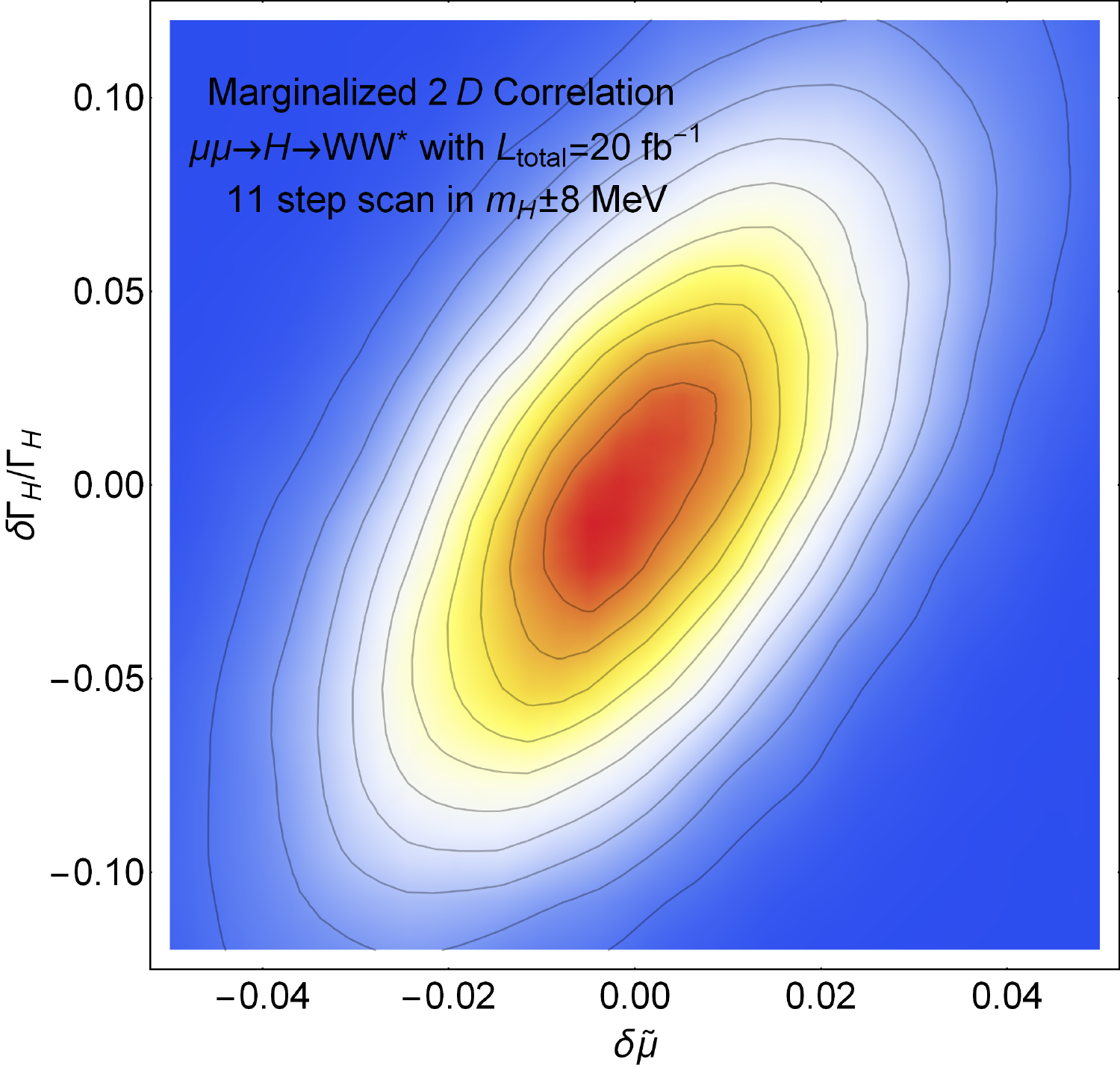}%{Rate_Width_correlation_WW.pdf}

\includegraphics[width=0.45\linewidth]{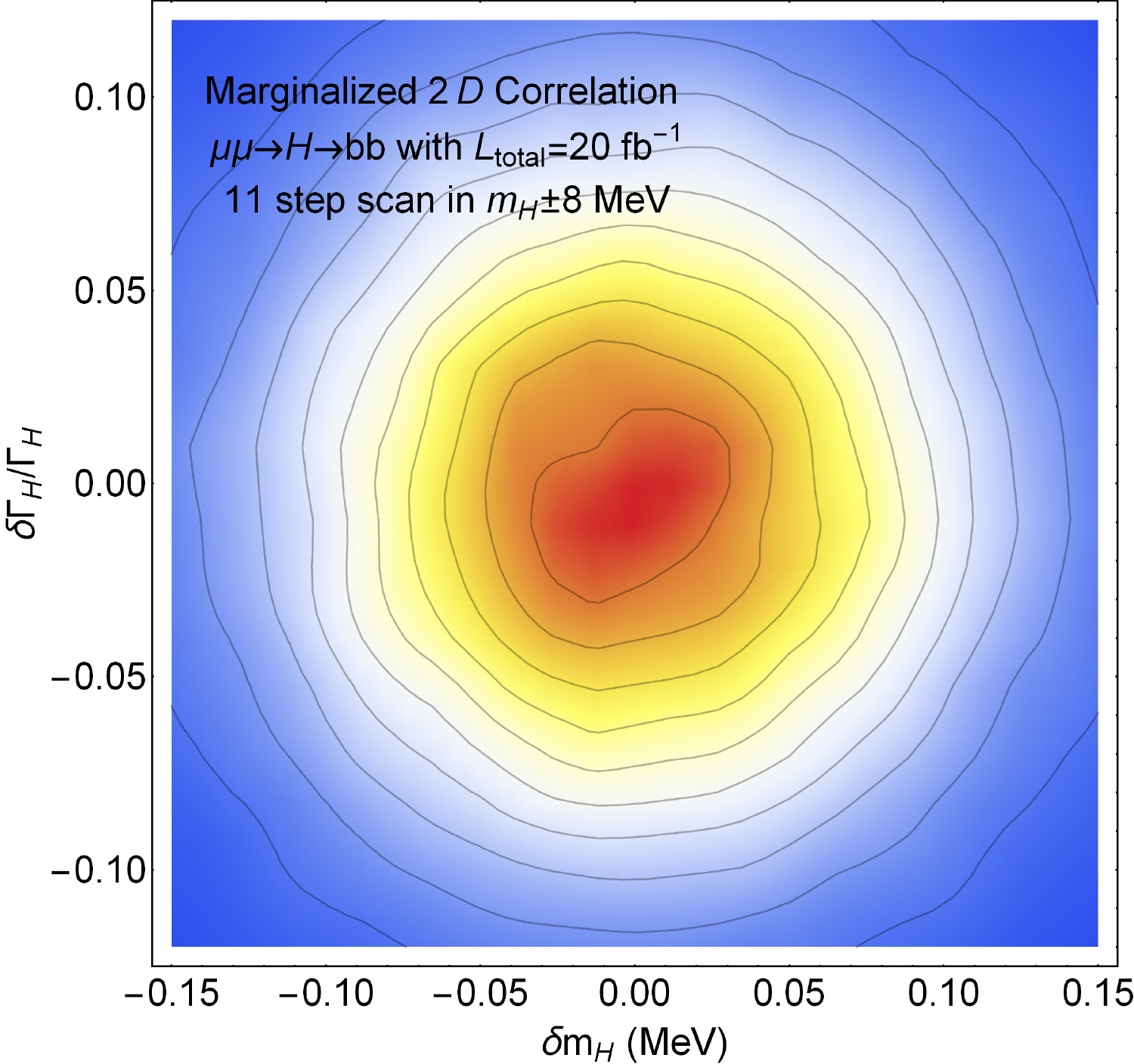}%{Mass_Width_correlation_b\bar b.pdf}
\includegraphics[width=0.45\linewidth]{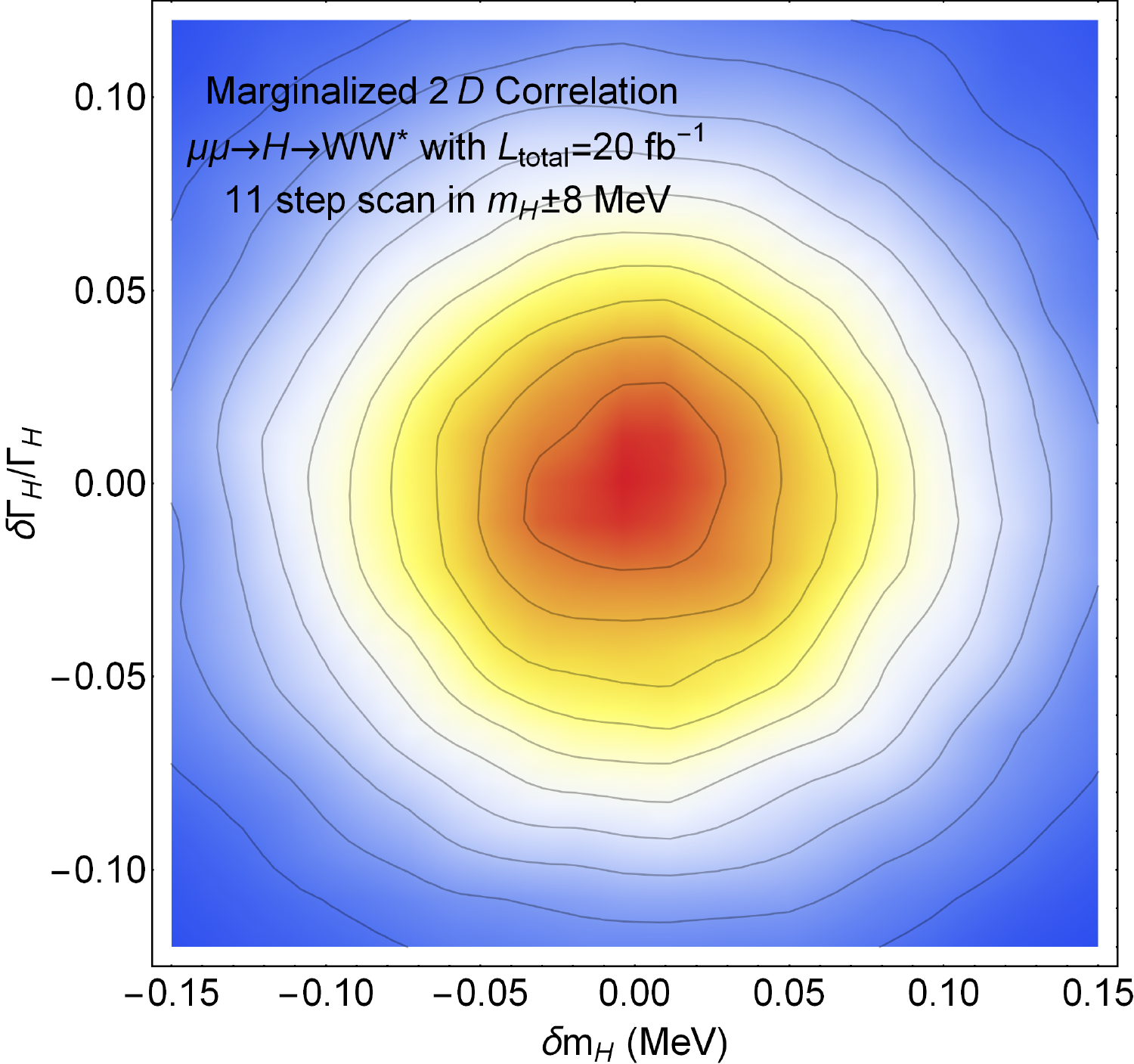}%{Mass_Width_correlation_WW.pdf}
\caption{The correlations between the different free parameters in the width fit adopted in this work. The corresponding numerical results are shown in \autoref{tab:widthresult}. Upper panels: the width and signal rate correlations ($\delta \Gamma_H/\Gamma_H$-$\delta\tilde\upmu$). Lower panels: the width and Higgs mass correlations ($\delta \Gamma_H/\Gamma_H$-$\delta m_H$). The results for the $b\bar b$ and $WW^*$ channels are shown in left and right panels, respectively.
\label{fig:fitcorrelations}}
\end{center}
\end{figure}

Our study also enables us to understand explicitly the {\it correlations} between the {\it fitting parameters $m_H$, $\Gamma_H$, and $\tilde \upmu$}. As anticipated and argued earlier, we expect a sizable correlation between the width and signal strength parameters. Their correlation are reported herein \autoref{tab:widthresult}. Again, we can see these correlations quantitatively and understand their impact, which is unknown prior to our study. We show correlations between different quantities after marginalizing over other parameters in \autoref{fig:fitcorrelations}. In this figure, the color represents the probability density. The contours are for equal probability density lines, with the left panels for the $b\bar b$ channel and the right panels for the $WW^*$ channel. The color coding and contours are mainly for reference purposes to visualize correlations and fluctuations of our simulations. In the upper two panels, we show the correlation between width precision and signal strength, where the $b\bar b$ channel is more strongly correlated compared with the $WW^*$ channel. In the lower two panels, we show the very small correlation between the Higgs width and Higgs mass precision. With this information and the Higgs exclusive channel precision projection in the next section, we can perform the global fit to reveal the Higgs physics potential at the 125 GeV muon collider.

{\it A note on the ISR effects in lineshape fitting:} Including ISR effects in the simulations and fitting procedure significantly increases the calculation time, as it involves one more layer of convolution. 
In this study we include both the ISR and BES effects. However, we also performed a version of our study without the ISR lineshape distortion effects for cross-check and for fast validation, only including an overall cross-section reduction. In hindsight, though perhaps not surprisingly, this does not change the results much. As we already noted in \autoref{sec:widthlinshape}, the ISR mainly reduces the rate but does not broaden the lineshape. This note might be a helpful input for full experimental simulations, where computational powers can be even more demanding and a limiting factor as detection effects are taken into account. 

%\clearpage

\section{Measurements}
\label{sec:meaasurements}

%=========================

To understand the full Higgs physics potential at the 125 GeV muon collider, one needs to study all primary Higgs decay channels that provide exclusive signals. These studies also impact the Higgs total width determination. This section describes our simulation and presents the projections for the expected precisions achievable on these exclusive channels.

%=========================

\subsection{Exclusive Higgs Rate Measurements}

The sensitivity study presented in this paper has been performed using {\tt MadGraph5\_aMC@NLO}~\cite{Frederix:2018nkq} to simulate both signal and background events in the different channels under consideration. Hadronic final states were further passed through {\tt Pythia8}~\cite{Sjostrand:2014zea}, and truth-level events were analysed with {\tt MadAnalysis5}~\cite{Conte:2012fm}. All simulations were performed at leading order, but signal cross-sections were properly scaled to include the effects of higher-order corrections, as as well as to reduce the rates resulting from considering the beam effects. In the {\tt MadGraph5\_aMC@NLO} generation of the samples we use the following basic cuts: $p_{T}^{j,b,\ell,\gamma}>5$ GeV, $\Delta R_{ij}>0.1$ and $|\eta_{i}|<2.44$, the latter to veto events within 10 degrees of the beam pipe and thus suppress the beam-induced backgrounds~\cite{Mokhov:2011zzd}. We use MLM matching~\cite{Mangano:2006rw} with a {\tt xqcut} value matching the cut on the $p_T$ of the jets. In the analysis of the events, we assume a perfect reconstruction of electrons and muons. For tau leptons, we assume a tagging efficiency $\varepsilon_{\tau,\mathrm{had}}=0.8$ and demand one tag for both fully hadronic and semi-leptonic tau decays. For the hadronic tagging, we adopt the values for the tag and mistag efficiencies detailed in Table~\ref{tab:bcgtag}. More details on the different backgrounds and the analyses are given below for each channel of interest.

In general, we follow a simple cut and count analysis to obtain the total rate for signal and backgrounds and obtain the corresponding sensitivity. In most of the 2-body final-state channels, we also explored the possibility of using the distribution of the production polar angle $\theta$ to help separate the Higgs signal from other SM backgrounds. Due to the scalar nature of the Higgs, the $\cos\theta$ distribution is flat for any two-particle final state of the signal. The background coming from an $s$-channel $Z/\gamma$ exchange (or $t/u$ channel muon exchange in the case of $\gamma\gamma$) exhibits different helicity structures and angular distributions. Since the muon beams are unpolarized and the polarization of the final state particles is generally difficult to measure (except taus), the electric charge of the final state particle offers a vital handle. If the charges are unknown, one can only measure the folded distribution, $|\!\cos\theta|$.   

\begin{figure}
\centering
\includegraphics[width=0.45\textwidth]{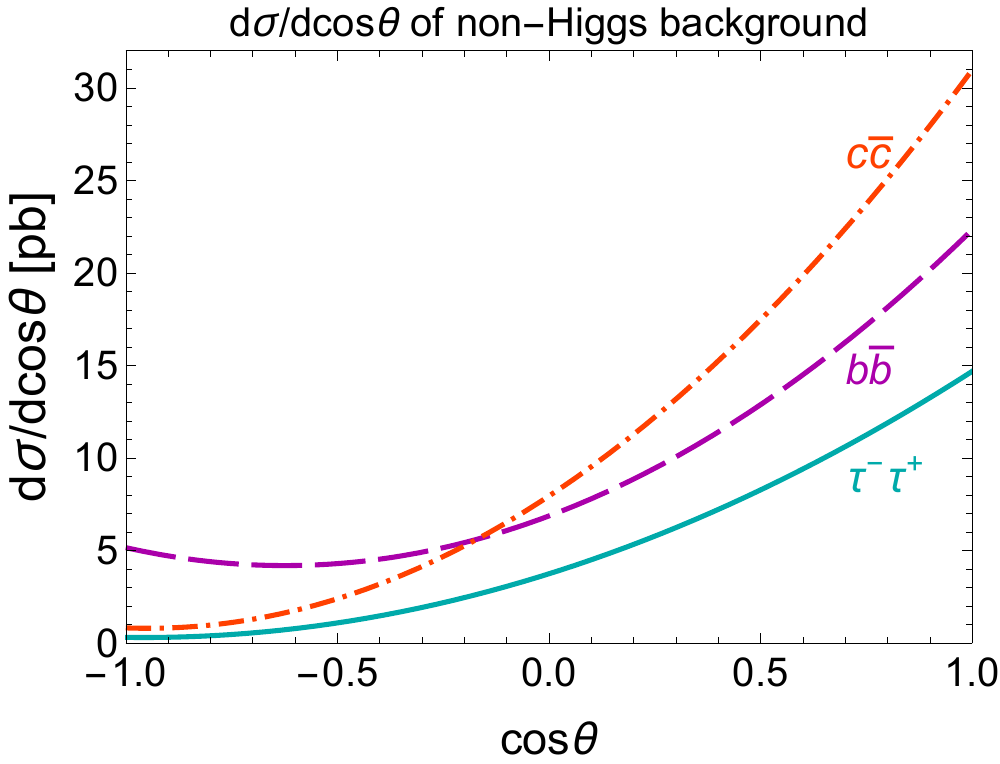} \hspace{0.1cm}
\includegraphics[width=0.45\textwidth]{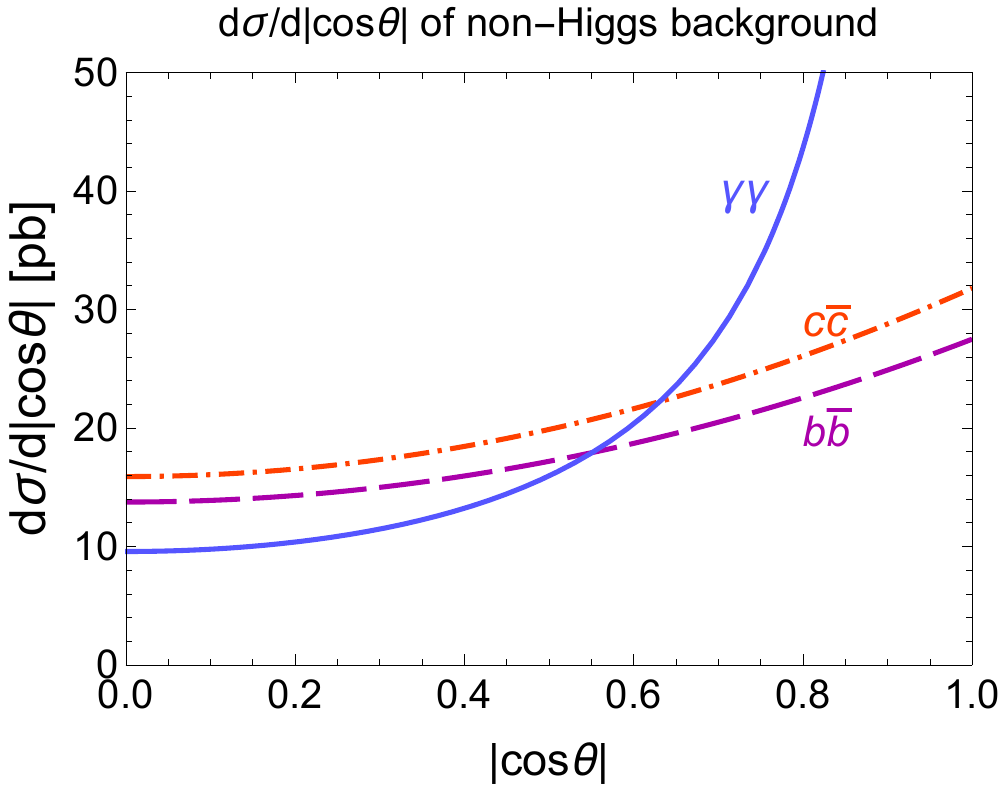} 
\caption{The differential cross-section for some of the non-Higgs backgrounds in terms of the production polar angle $\theta$.  On the left panel, we assume the two final state particles can be distinguished and the sign of $\cos\theta$ is measured.  On the right panel, only the folded distribution in terms of  $|\!\cos\theta|$ is measured.}
\label{fig:ddcos}
\end{figure}

\autoref{fig:ddcos} shows the differential cross-section for some of the non-Higgs backgrounds. On the left panel, the results for $b\bar{b}$, $c\bar{c}$ and $\tau^-\tau^+$ are shown, assuming the two final state fermions can be distinguished.  The angle $\theta$ is always defined to be between the initial $\mu^-$ and the $b/c/\tau^-$. This requires the electric charge of at least one of the final state fermions to be measured. While the forward-back asymmetries are generally suppressed at the $Z$-pole (which is due to the coincidence $g_L \approx - g_R$ for leptons), for $\sqrt{s}=125$\,GeV they are quite large as seen in \autoref{fig:ddcos}.  The $\cos\theta$ distribution thus offers a useful handle for separating backgrounds from the Higgs signal, as the latter has a flat distribution. For $b\bar{b}$ and $c\bar{c}$, the charge is not always measured, in which case one could only measure the folded differential cross-section, $d\sigma/d|\!\cos\theta|$, as shown on the right panel of \autoref{fig:ddcos}.  The folded distribution is not very powerful in terms of the signal-background separation. Similarly, for the $\gamma\gamma$ final states, one is also only able to measure $d\sigma/d|\!\cos\theta|$.  Due to the $t/u$-channel muon exchange, the $\gamma\gamma$ background dominates the forward region.

In what follows, we summarize some details of the simulations used for the main decay channels considered in this study.

\begin{itemize}

\item {\bf $b\bar{b}/c\bar{c}/gg$:} The performances in the $b\bar{b}/c\bar{c}/gg$ channels depend heavily on the flavor tagging efficiencies.  We assume the muon collider could achieve a similar performance as $e^+e^-$ colliders and extract the tagging efficiencies from existing CEPC studies. In particular, we chose the $b$- and $c$-tagging efficiencies to be 0.7 and 0.6, respectively, and extracted the corresponding mistag rates. The efficiencies for a jet to be tagged as a gluon (or, more precisely, to escape all flavor tags) are not available. 
We consider a scenario where a jet is tagged as either $b$, $c$, or $g$, such that the ``gluon-tagging'' efficiency can be calculated from the b and c efficiencies. 
The tagging-efficiencies are shown in table~\ref{tab:bcgtag}.  The main background comes from $s$-channel $Z/\gamma$ production with $b\bar{b}$, $c\bar{c}$ and light quark final states.  We assume the light quarks ($u,d,s$) are indistinguishable from gluons, and the gluon-tagging efficiencies also apply to them. We perform a global fit to all six channels with different tags (including the three channels with mixed tags) with both signal and backgrounds to extract the bound on the signal rates of the actual $b\bar{b}$,$c\bar{c}$, $gg$ channels. 

In addition, we make use of the $\cos\theta$ distribution in \autoref{fig:ddcos} to try to obtain a better signal-background separation. In general, only the folded distribution is available.  This also applies to the $gg$ channel where the light quark backgrounds exhibit the same $|\!\cos\theta|$ distributions as $b\bar{b}$ and $c\bar{c}$.  For the $b\bar{b}$ and $c\bar{c}$ channels, we assume the charge can be measured if 1) both flavor tags are correct and 2) at least one of the $b$ (or $c$) decays leptonically. 
We found that our crudely binned distributions provide only a minor improvement in the precision reach at the 3\% level. 

\begin{table}%[h]
\centering
\begin{tabular}{|c||c|c|c|} \hline
 &  \multicolumn{3}{|c|}{actual} \\  \hline
&   $b$  &  $c$  &  $g$ \\ \hline\hline
tagged as $b$ & 0.7 & 0.04 & 0.004  \\  \hline
tagged as $c$ & 0.2 & 0.6 & 0.07  \\  \hline
tagged as $g$ & 0.1 & 0.36 & 0.926 \\  \hline
\end{tabular}
\caption{The probability of an actual $b$, $c$, or $g$ to be tagged as $b$, $c$ or $g$.  The ``gluon-tagging'' efficiency in the last row is calculated by assuming that a particle is tagged as either $b$, $c$, or $g$ (no double counting), so that the 3 numbers in each column add up to 1. 
}
\label{tab:bcgtag}
\end{table}

\item {\bf $\tau^+\tau^-$:}  
For the tau tagging efficiencies, a reference benchmark can be found in the CLIC report~\cite{Abramowicz:2016zbo}, which keeps only the hadronic taus and assumes $73\%$ of them being tagged.  
As mentioned above, we will assume a slightly more optimistic hadronic tau tagging efficiency of $80\%$ and require at least one  
tau to be tagged for the signal selection, assuming its charge is always measured. We consider only the tau pairs from $s$-channel $Z/\gamma$ exchanges for the background. Other background comes from quarks being mistagged as taus and is under control for a sufficiently low mistag rate. 
To use $\cos\theta$ distribution, we perform a $\chi^2$ fit to the binned $\cos\theta$ distribution with both signal and background to extract the precision reach on the signal rate. The distribution is divided into 20 bins, and the rate measurements of different bins are assumed to be uncorrelated. The binned analysis significantly improves the one with only the total rate, with an improvement in the precision of approximately 30$\%$. 
The precision can, in principle, be further improved by exploiting the measurements of tau final state polarizations.  
It is also possible to tag the leptonic decay modes of the tau by exploiting track information.
The CEPC analysis~\cite{Yu:2020bxh} reports an overall tagging efficiency of around $80\%$ including all tau decay channels.   
We leave a more optimized $\tau^+\tau^-$ analysis to future studies.

\item {\bf $\gamma\gamma$:} Since the background in this channel is concentrated in the forward region, the binned analysis can provide a significant improvement. We thus perform a binned distribution analysis similar to the one of $\tau^+\tau^-$, but with the $|\!\cos\theta|$ distribution. Although we see
an improvement by a factor of $\sim$33\% with respect to the cut and count 
precision, the signal rate is too low for this channel to be of practical use. 

\item {\bf $4f$:} We study the different four fermion final states, $H\to V V^* \to 4f$ separately. 
One must note that while certain channels such as $H\to 4\ell$ or $H\to \ell \nu jj$ receive only contributions from neutral (e.g. $ZZ^*$) and charged (e.g. $WW^*$) current electroweak interactions, respectively, others final states, such as $H\to e^+ e^- \nu_e \bar{\nu_e}$, are produced via both mechanisms, including their interference, though in these cases the $WW^*$ contributions dominates. In these cases, to obtain a certain separation between the $ZZ^*$ and $WW^*$ contributions, we apply a cut on the difermion invariant mass, e.g., $M_{ee}\in[80,100]$ GeV, selecting such events for the neutral current and the remainder for the charged current process. 
This procedure is applied to the leptonic final states. 
For all (semi) leptonic final states, we consider the corresponding electroweak backgrounds from $\mu^+ \mu^- \to Z Z^*, W W^*$ 
and apply the above-mentioned cuts whenever we aim to separate neutral and charged current contributions.
For the $H\to 4j$ channel, apart from the corresponding electroweak backgrounds from $\mu^+ \mu^- \to Z Z^*, W W^*$, we also include the contributions from $\mu^+ \mu^- \to jj, jjj$. 
Jets are reconstructed using the anti-$k_t$ algorithm~\cite{Cacciari:2008gp}, using a minimum $p_T$ of 5 GeV and a radius of 0.5. 
To separate signal and background we exclude reconstructed events with $N_j<4$, and also reject events with soft jets $p_{T,j_3}<10$ GeV, as well as those with $M_{j_1,j_2}>85$ GeV.
Unfortunately, for this fully hadronic decay, $H\to 4j$ decay, the final state interactions deform the invariant mass shapes, making the previous simple cut and count procedure to distinguish between neutral and charged current contributions not possible. Hence, we report the result in a separate row in \autoref{tab:sensitivity}.
Finally, for all the purely neutral current channels, we also impose a lower cut in the difermion invariant masses $M_{ff}>5$ GeV to remove contributions from on-shell photons. 

\end{itemize}

%=========================

\subsection{Results}

In \autoref{tab:sensitivity}, we show the estimation for the statistical precision of different exclusive Higgs channels at the 125 GeV muon collider, for two benchmarks of integrated luminosities of $5\fbi$ and  $20\fbi$. 
This table is based upon our study, with a simple cut-and-count analysis, which has room for improvement.
We also quote the expected improved precision in those channels where we can use the information from the polar angle differential distributions to enhance the sensitivity to the signal, though, other than for the $\tau^+ \tau^-$ channel, this proves to be of limited use. Finally, we do not show $4f$ channels in the table that contribute in a negligible manner to the overall precision of the $H\to WW^*, ZZ^*$ decays. 

As anticipated, the muon collider Higgs factory has good cross-section measurement for Higgs decay channels such as $b\bar b$ but is deficient on statistically limited channels such as $\gamma\gamma$. However, we note that the $H\to WW^*$ and $H\to ZZ^*$ modes can be measured to a few percent. This advantage comes from the better signal to background ratio at muon colliders than other electron-positron Higgs factories, owing to a lower center-of-mass-energy and lack of combinatorics from multi-jet final states. Such an observation also leads to a lower background expected for other Beyond the SM (BSM) Higgs programs, in particular, for Higgs exotic decays in the fully hadronic channel~\cite{Liu:2016zki}.

We also want to comment on the impact of beam-induced-background (BIB)~\cite{BIBref} for our results. The BIB considerations mainly affect our results from two perspectives. First, shielding BIB limits the signal acceptance angle from the detector design point of view. In our study, we veto events within 10 degrees of the beam pipe. Should we veto further, one can rescale the signal rate accordingly, as it is a spinless particle decay. Second, the BIB would populate a lot of low energy activities. Our selection for Higgs decays focus on high energy decay products, more than 5 GeV in $p_T$ and it should be robust against BIB. Still, it would be highly needed to verify these understanding via a full simulation.

\begin{table}[thbp]
  \centering
    {%\small
      \begin{tabular}{cccccc c } 
        \ctoprule
        Channel &
        Rate &
        Signal &
        Background &
        \multicolumn{3}{c}{Precision [$\%$]} \\
        $\mu^+ \mu^- \to h \to X$ &
        [pb] &
        Events &
        Events &
        \multicolumn{2}{c}{{\scriptsize Cut \& Count}} & {\scriptsize Binned} \\ [0.1cm]
         & &
        \multicolumn{5}{c}{Results for 5/20 fb$^{-1}$} \\ [0.1cm] 
        \cmrule
        &&&&&&\\ [-0.25cm]
        $b\bar{b}$ &
        13 &
        19000/77000 &
        45000/180000 &
        1.0/0.51 & 
        & 0.97/0.49 \\ [0.1cm]
        %\crowcolor 
        $c\bar{c}$ &
        0.63 &
        2300/9200 &
        43000/170000 &
        24/12 & 
        & 23/12 \\ [0.1cm]
        $gg$ &
        1.8 &
        5400/22000 &
        260000/10$^6$ &
        11/5.5 & 
        & 11/5.3 \\ [0.1cm]
        \cmrule
%-----------------------------------
        &&&&&\\ [-0.35cm]
        %\crowcolor 
        $\tau_{\rm had}^+ \tau_{\rm had}^-$ &
        0.58 &
        1400/5600 &
        19000/76000 &
        10/5.1 & \multirow{2}{*}{6.8/3.4} & \multirow{2}{*}{4.8/2.4}\\ [0.1cm]
        $\tau_{\rm had}^+ \tau_{\rm lept}^-$  &
        0.63 &
        1500/6100 &
        18000/71000 &
        9.1/4.5 & & \\ [0.1cm]
        \cmrule        
%----------------------------------- 
        &&&&&&\\ [-0.35cm]
        $\gamma\gamma $ &
        0.05 &
        150/605 &
        180000/730000 &
        280/140 &  
        & 190/94 \\ [0.1cm]
        \cmrule
%----------------------------------- 
        &&&&&\\ [-0.35cm]
% 4f channels: NC
         $2\ell 2 q~(\ell=e,\mu)$ &
         0.05&
         130/530&
         1200/4800&
         28/14 & \multirow{4}{*}{5.8/2.9}
         &\\ [0.1cm]
         $2\nu 2j  $ &
         0.16&
         450/1800&
         320/1300&
         6.1/3.1 & 
         &\\ [0.1cm]
        $2e 2\nu^\ddag $ &
         0.005&
         8/33&
         0/1&
         35/18 &   
         &\\ [0.1cm]
        $2\mu 2\nu^\ddag $ &
         0.005&
         9/35&
         0/1&
         34/17 &
         &\\ [0.1cm]
        \cmrule
%----------------------------------- 
        &&&&\\ [-0.35cm]
% 4f channels: CC
        $e\nu\mu\nu$ &
         0.11 &
         320/1300&
         9/35 & 
         5.7/2.8 & 
        &\\ [0.1cm]
        $\ell\nu\tau_{\rm had}\nu~(\ell=e,\mu)$ &
        0.14 &
         330/1300 &
         8/32 & 
         5.6/2.8 &
        &\\ [0.1cm]
        $\ell\nu jj~(\ell=e,\mu)$ &
        1.4 &
        3800/15000 &
        88/350 & 
        1.6/0.82 & 
        &\\ [0.1cm]
        $\tau_{\rm had}\nu jj~$ &
        0.45 &
        1000/4000 &
        20/79 & 
        3.2/1.6 &
        1.3/0.67 % Common unc.  
        &
        \\ [0.1cm]
        $2e 2\nu^\dagger $ &
         0.06&
         160/660&
         86/340&
         9.6/4.8 &
         &\\ [0.1cm]
         $2\mu 2\nu^\dagger $ &
         0.06&
         160/650&
         76/310&
         9.5/4.7 &
         &\\ [0.1cm]
        $2\tau_{\rm had} 2\nu^\dagger $ &
         0.023&
         46/180&
         24/97&
         18/9.1 &
         &\\ [0.1cm]
        \cmrule
%----------------------------------- 
        &&&&&&\\ [-0.35cm]
% 4f channels: CC+NC
        $4j (j\not = b)$ &
         2.3&
         3400/14000&
         51000/210000&
         6.8/3.4 &
         &\\ [0.1cm]
        \cbottomrule
      \end{tabular}
    }
\caption{Sensitivity to each of the Higgs decay channels at a 125~GeV muon collider. The signal and background quoted in this table correspond to total events expected for a luminosity of 5 and 20 fb$^{-1}$ with our scan strategy. The second column from the ``Cut \& count'' method shows the combined precision from different channels (e.g., $\tau^+ \tau^-$), assuming a common signal strength.
    The last column shows the precision exploiting the information in the polar angle distributions for the two-body final states.
    For the $2\ell 2\nu$ channels, we show separately the sensitivities obtained after splitting the samples using the $M_{\ell\ell}$ cut around the $Z$ mass, collecting the sample of events passing that cut in the $2\ell 2\nu^\ddag$ ``category'' in the table and the remainder in the $2\ell 2\nu^\dagger$ one. 
    In the second column, we also report the on-shell cross-section, which multiplies the effective factor in  \autoref{eq:effectivexs} from our scan to provide the effective cross-section in \autoref{eq:onshellxs}.
    \footnote{Note that the lineshape scan channels $b\bar b$ and $W W^*$ correlates with the Higgs width parameter directly, as discussed in \autoref{sec:widthresults} and reported in \autoref{tab:widthresult}. The determined exclusive precision is worse than reported here in this table with such correlation. In other words, the exclusive precision reported in these two channels is determined in a particular direction in multi-parameter space, and its information has already been fully taken into account in the width fit so we shall not treat them as new inputs beyond \autoref{tab:widthresult}.}
    \label{tab:sensitivity}
    }
\end{table}

\clearpage

%=============================================================

\section{Higgs Coupling Interpretation}
\label{sec:couplings}

In this section we carry out a Higgs precision global fit with the projected sensitivities in the Higgs lineshape scan and the exclusive Higgs measurements. We describe our frameworks and discuss the results in what follows.

%=========================

\subsection{The $\kappa$ Framework}

With the precision of the measurement of the Higgs width and the correlations in \autoref{tab:widthresult}, and the exclusive measurements in \autoref{tab:sensitivity}, one can perform a global fit to derive the expected precision on the Higgs couplings at a 125\,GeV muon collider.~\footnote{The $\kappa$ fits presented in this section were performed using the {\tt HEPfit} code~\cite{DeBlas:2019ehy} and following a Bayesian approach, as in~\cite{deBlas:2019rxi}. The uncertainties reported are defined as the square root of the variances for the different parameters, obtained from the posterior predictive of the fits.}
For that purpose, we use in this section the so-called $\kappa$ framework. The cross-sections are decomposed as discussed in \autoref{sec:width}, and the effective cross-section treatment provides a good approximation to the threshold scan measurement benchmark we implement in this study. These effective cross sections are thus parameterized in terms of scaling parameters $\kappa$, interpreted as coupling modifiers:
\begin{equation}
( \sigma^{\rm eff} )_{\mu^+ \mu^- \to H \to f} \propto \frac{\Gamma_{\mu^+ \mu^-} \Gamma_f}{\Gamma_H} = \frac{ \Gamma_{\mu^+ \mu^-}^{\SM} \Gamma_{f}^{\SM} }{\Gamma_H^{\SM}}\times\frac{\kappa_\mu^2 \kappa_f^2}{ \kappa_\Gamma} (1-{\rm BR}_{\rm exo}) \quad ,
\label{eq:kappa_zw}
\end{equation}
where we have also assumed the possibility of Higgs decaying into {\em exotic} ({\it i.e.} non-SM) final states, parameterized by ${\rm BR}_{\rm exo}\geq 0$, so that the total width is expressed as $\Gamma_H = \frac{ \Gamma_{H}^\text{SM} \cdot \kappa_\Gamma }{ 1-{\rm BR}_{\rm exo} }$ with
$\kappa_\Gamma \equiv \sum\limits_{j} \kappa_j^2 \Gamma_{j}^\text{SM} /
\Gamma_{H}^\text{SM}\, .$  
The direct width measurement at the 125\,GeV muon collider allows closing a fit where the Higgs width is a free parameter.  
In the $\kappa$ fits performed in this paper, we take the following Higgs coupling modifiers as free parameters in the fits:
$$\left\{\kappa_Z, \kappa_W, \kappa_t,~\kappa_c,~\kappa_b,~\kappa_\tau,~\kappa_\mu, \kappa_g,~\kappa_\gamma,~\kappa_{Z\gamma}\right\}$$
where the last three parameters refer to effective coupling modifiers for the SM loop-induced processes, to parameterize the possible presence of extra particles in the loops.
Together with ${\rm BR}_{\rm exo}$, this makes a total of 11 free parameters. 
Any other coupling modifier is set to its SM value $\kappa_i=1$.\footnote{For the HL-LHC fit, because of the lack of projections to directly constrain the charm coupling, we also set $\kappa_c=1$.}
We will also consider a scenario where the Higgs boson does not have any exotic decay (${\rm BR}_{\rm exo}=0$), and the Higgs width is given by the sum of the widths to all SM decay products, $\Gamma_H =\Gamma_{H}^\text{SM} \cdot \kappa_\Gamma$. This is referred to later as the {\em constrained} fit.
In both the {\em standard} and the constrained fits, we assume theory calculations will achieve the required level of precision to match the experimental one at each collider. Thus, we neglect any possible uncertainty associated with missing higher-order corrections to the SM processes. We do, however, consider the projected parametric uncertainties due to the projected experimental knowledge of the SM input parameters, as in \cite{deBlas:2019rxi}. In addition, for the muon collider, we consider the expected below per-mille precision in $m_H$, which implies that any associated uncertainty to this parameter can be effectively ignored.

%---------------------------

Our results for the global Higgs-coupling fit in the $\kappa$ framework are shown in \autoref{fig:kapparesult} for the 125\,GeV muon collider, assuming a luminosity of 20 fb$^{-1}$. 
The alternative muon Higgs factory scenario with 5 fb$^{-1}$ luminosity is shown in \autoref{fig:kapparesult5ifb}.
The numerical results of all our fits are provided in \autoref{t:Kappa_sensitivity} (for the constrained fit) and \autoref{t:Kappa_sensitivity_2} (for the standard fit, with BR$_{\rm exo}\geq 0$ a free parameter, i.e., allowing Higgs decays into new particles in $\Gamma_H$).
Comparisons are made with several other future collider scenarios, including the HL-LHC and a potential $e^+e^-$ collider running at 240\,GeV.  For Higgs measurements at HL-LHC, we use the inputs of the S2 scenario in Ref.~\cite{Cepeda:2019klc}.  For the $e^+e^-$ collider, we use the CEPC projections~\cite{An:2018dwb,CEPCStudyGroup:2018ghi}, while similar reaches are also obtained by FCC-ee 240\,GeV\cite{FCC:2018evy} and ILC 250\,GeV\cite{Bambade:2019fyw}.  
The four columns in \autoref{fig:kapparesult} correspond to the HL-LHC S2, a 240\,GeV $e^+e^-$ collider, a 125\,GeV muon collider ($20\ifb$) and the combination of the $e^+e^-$ and the muon colliders, respectively.  For the lepton collider scenarios, we also combine the measurements with the HL-LHC ones, assuming the latter would be available by the time the lepton colliders are running. For the constrained fits, the results are shown by the horizontal marks.

\begin{figure}[t]
\centering
\includegraphics[width=\linewidth]{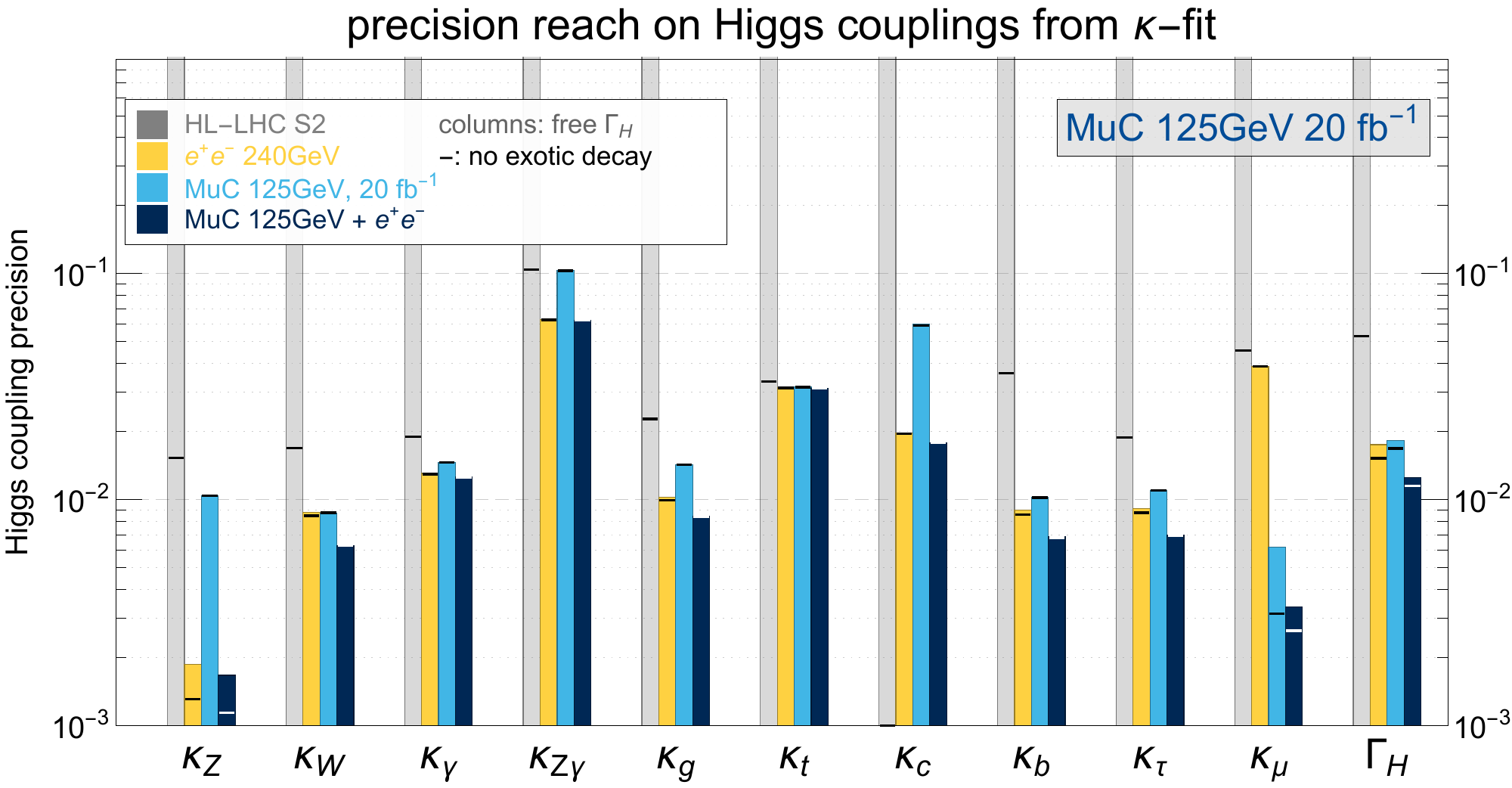}
\caption{
The Higgs coupling precision from a global fit of Higgs measurements in the $\kappa$-framework. The four columns represent the HL-LHC S2 scenario, a circular $e^+e^-$ collider at $240\,$GeV, a muon collider at 125\,GeV with a total integrated luminosity of $20\ifb$, and the combination of the $e^+e^-$ and the muon collider, respectively. The measurements are combined with the HL-LHC S2 for all the lepton collider scenarios. The column shows results with $\Gamma_H$ treated as a free parameter; the horizontal marks show the ones assuming that the Higgs has no exotic decay. 
}
\label{fig:kapparesult}
\end{figure}
\begin{figure}[h]
\centering
\includegraphics[width=\linewidth]{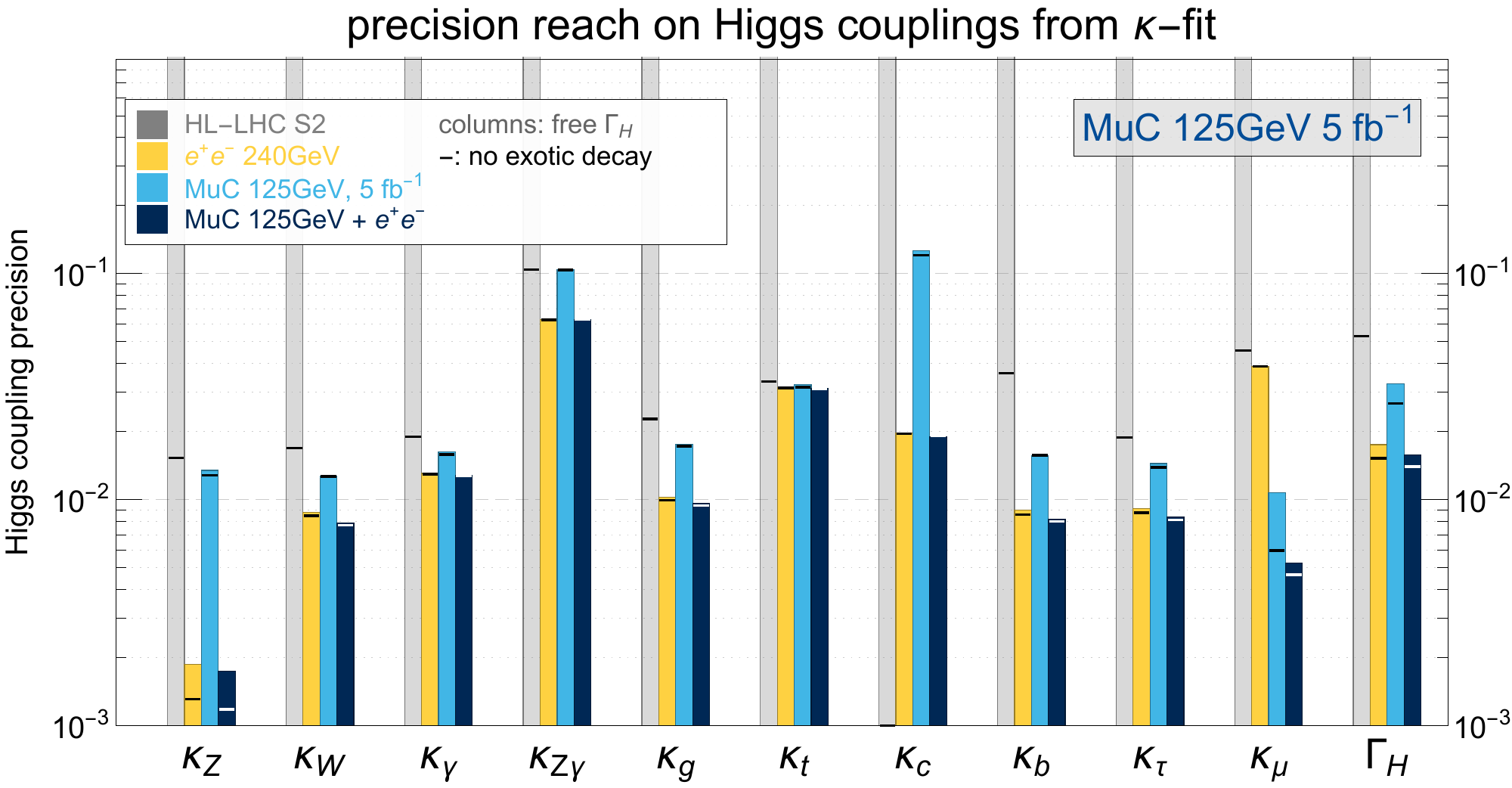}
\caption{
The same as \autoref{fig:kapparesult} but using the luminosity benchmark of 5 fb$^{-1}$ for the muon collider.
}
\label{fig:kapparesult5ifb}
\end{figure}

Focusing on the 20 fb$^{-1}$ benchmark scenario, we observe that, overall, the 125\,GeV muon collider significantly improves the Higgs coupling constraints of the HL-LHC measurements. This is obviously true in the case with a free Higgs width, where the LHC measurements alone are unable to close the fit. In the constrained fit, sizable improvements are also observed in many of the Higgs couplings as well as the Higgs total width. In particular, it is challenging to probe the charm Yukawa coupling at the LHC, while a precision of $\sim 6\%$ could be achieved at the 125\,GeV muon collider. Not surprisingly, the muon Yukawa coupling precision also gets a significant boost (of more than one order of magnitude) at the 125\,GeV muon collider, thanks to its unique $\mu^+\mu^- \to H$ production channel. Even a future $e^+e^-$ collider could not compete with the 125\,GeV muon collider on the muon Yukawa coupling determination. On the other hand, a 240\,GeV $e^+e^-$ collider is much better at measuring the $HZZ$ coupling, owing to its excellent inclusive $ZH$ production measurement. For the other couplings, the two colliders' reaches are comparable when individually combined with HL-LHC, despite the fact-of-five difference in the total number of Higgs events. For $\kappa_\gamma$,  $\kappa_{Z\gamma}$ and $\kappa_t$, this is mainly due to the constraints from HL-LHC; for the other couplings, it is indeed a consequence of clean measurement, (in many cases) smaller backgrounds, and the ability to directly measure the width, as we emphasized in the previous section. 
The comparison discussion changes for the 5 fb$^{-1}$ scenario. Even with the reduced statistics, leading to a factor of two deterioration in the (standalone) muon collider precision, the muon coupling could still be known at the percent level and, thus, better than any other collider. All the other parameters are more than a factor of two less precise in terms of precision compared to a $e^+ e^-$ Higgs factory due to a factor of 20 fewer statistics for the 5 fb$^{-1}$ scenario.

The complementarity between a 125\,GeV muon collider and a 240\,GeV $e^+e^-$ collider is also interesting and highly non-trivial.  First, we note that the two colliders have similar reaches on the Higgs total width while using completely different methods (1.8\% at $20~\fbi$ from a direct threshold scan\footnote{The reasons for a better total width precision at muon collider compared to shown in \autoref{tab:widthresult} is that the HL-LHC inputs reduces the correlation between width and signal strength parameter $\tilde \upmu$.} {\it vs.} 1.8\%  for the determination obtained combining the inclusive $ZH$ cross-section and the individual rates). They thus provide important consistency checks on the properties of the Higgs boson, which is not reflected in the numerical results. The combination further improves the overall reach to $1.3\%$. Many Higgs couplings also receive sizable improvements from the combination. For both the 5 and 20 fb$^{-1}$ benchmarks, the improvement turns out to be rather significant for $\kappa_\mu$ (compared with the muon-collider-alone result), while one would naively expect it to be completely dominated by the muon collider measurements. This is because these two colliders, with different production channels, have quite different correlation matrices for the Higgs couplings. The $e^+e^-$ collider measurements could help resolve the large correlation between $\kappa_\mu$ and $\kappa_b$ in the muon collider measurements.
%

%======================================================
%
\begin{table}[t]
  \centering
    \vspace{0.25cm}
    {\small
      \begin{tabular}{c|c|c|cc|cc} 
        \ctoprule
        %\crowcolor
        &
        HLLHC &
        HLLHC &
        \multicolumn{2}{c|}{HLLHC} &
        \multicolumn{2}{c}{HLLHC}   \\
         &
         &
         
        + $e^+e^-$ $H$ fact.  &
        
        \multicolumn{2}{c|}{+ $\mu$-coll.} &

        \multicolumn{2}{c}{+ $\mu$-coll.}  \\
        
        Coupling &
         &
         
          &
        
        \multicolumn{2}{c|}{} &

        \multicolumn{2}{c}{+$e^+e^-$ $H$ fact.}  \\
        
        &&&5 fb$^{-1}$&20 fb$^{-1}$&5 fb$^{-1}$&20 fb$^{-1}$\\
        
        \cmrule
        &&&&&&\\ [-0.25cm]
%-----------------------------------
$\kappa_{W}~[\%]$   & $1.7$   & $0.85$   & $1.3$   & $0.88$   & $0.77$   & $0.62$  \\ [0.1cm]
$\kappa_{Z}~[\%]$   & $1.5$   & $0.13$   & $1.3$   & $1.0$   & $0.12$   & $0.11$  \\ [0.1cm]
$\kappa_{g}~[\%]$   & $2.3$   & $0.99$   & $1.7$   & $1.4$   & $0.94$   & $0.84$  \\ [0.1cm]
$\kappa_{ \gamma}~[\%]$   & $1.9$   & $1.3$   & $1.6$   & $1.5$   & $1.3$   & $1.3$  \\ [0.1cm]
$\kappa_{Z \gamma}~[\%]$   & $10$   & $6.2$   & $10$   & $10$   & $6.2$   & $6.2$  \\ [0.1cm]
$\kappa_{c}~[\%]$   & $-$   & $2.0$   & $12.$   & $5.9$   & $1.9$   & $1.8$  \\ [0.1cm]
$\kappa_{t}~[\%]$   & $3.3$   & $3.1$   & $3.1$   & $3.1$   & $3.1$   & $3.1$  \\ [0.1cm]
$\kappa_{b}~[\%]$   & $3.6$   & $0.86$   & $1.6$   & $1.0$   & $0.8$   & $0.68$  \\ [0.1cm]
$\kappa_{\mu}~[\%]$   & $4.6$   & $3.9$   & $0.59$   & $0.31$   & $0.46$   & $0.26$  \\ [0.1cm]
$\kappa_{\tau}~[\%]$   & $1.9$   & $0.87$   & $1.4$   & $1.1$   & $0.81$   & $0.69$  \\ [0.1cm]
\cmrule
$\Gamma_{H}~[\%]$   & $5.3$   & $1.5$   & $2.7$   & $1.7$   & $1.4$   & $1.1$  \\ [0.1cm]
        \cbottomrule
      \end{tabular}
    }
\caption{Results from the $\kappa$ fit assuming no BSM contributions to the Higgs width.
    The results with a future muon collider correspond to scenarios where the total luminosities are assumed to be 5 and 20 fb$^{-1}$. 
    The last row also includes the derived precision on the total Higgs width from the fits.
    \label{t:Kappa_sensitivity}
    }
\end{table}
\begin{table}[t]
  \centering
    \vspace{0.25cm}
    {\small
      \begin{tabular}{c|c|c|cc|cc} 
        \ctoprule
        %\crowcolor
        &
        HLLHC &
        HLLHC &
        \multicolumn{2}{c|}{HLLHC} &
        \multicolumn{2}{c}{HLLHC}   \\
         &
         &
         
        + $e^+e^-$ $H$ fact.  &
        
        \multicolumn{2}{c|}{+ $\mu$-coll.} &

        \multicolumn{2}{c}{+ $\mu$-coll.}  \\
        
        Coupling &
         &
         
          &
        
        \multicolumn{2}{c|}{} &

        \multicolumn{2}{c}{+$e^+e^-$ $H$ fact.}  \\
        
        &&&5 fb$^{-1}$&20 fb$^{-1}$&5 fb$^{-1}$&20 fb$^{-1}$\\
        
        \cmrule
        &&&&&&\\ [-0.25cm]
%-----------------------------------
$\kappa_{W}~[\%]$   & $-$   & $0.88$   & $1.3$   & $0.88$   & $0.79$   & $0.63$  \\ [0.1cm]
$\kappa_{Z}~[\%]$   & $-$   & $0.19$   & $1.3$   & $1.0$   & $0.17$   & $0.17$  \\ [0.1cm]
$\kappa_{g}~[\%]$   & $-$   & $1.0$   & $1.7$   & $1.4$   & $0.96$   & $0.84$  \\ [0.1cm]
$\kappa_{ \gamma}~[\%]$   & $-$   & $1.3$   & $1.6$   & $1.5$   & $1.3$   & $1.3$  \\ [0.1cm]
$\kappa_{Z \gamma}~[\%]$   & $-$   & $6.3$   & $10$   & $10$   & $6.3$   & $6.2$  \\ [0.1cm]
$\kappa_{c}~[\%]$   & $-$   & $2.0$   & $13$   & $6.0$   & $1.9$   & $1.8$  \\ [0.1cm]
$\kappa_{t}~[\%]$   & $-$   & $3.1$   & $3.2$   & $3.1$   & $3.1$   & $3.1$  \\ [0.1cm]
$\kappa_{b}~[\%]$   & $-$   & $0.9$   & $1.6$   & $1.0$   & $0.82$   & $0.69$  \\ [0.1cm]
$\kappa_{\mu}~[\%]$   & $-$   & $3.9$   & $1.1$   & $0.62$   & $0.53$   & $0.34$  \\ [0.1cm]
$\kappa_{\tau}~[\%]$   & $-$   & $0.91$   & $1.5$   & $1.1$   & $0.84$   & $0.7$  \\ [0.1cm]
$\mathrm{BR}_{\mathrm{exo}}^{95\%}<$   & $-$   & $1.1$   & $4.7$   & $3.0$   & $1.1$   & $1.0$  \\ [0.1cm]
\cmrule
$\Gamma_{H}~[\%]$   & $-$   & $1.7$   & $3.3$   & $1.8$   & $1.6$   & $1.2$  \\ [0.1cm]

        \cbottomrule
      \end{tabular}
    }
\caption{Results from the $\kappa$ fit including the possibility of Higgs decays into light BSM degrees of freedom, parameterized by BR$_{\rm exo}$. The results with a future muon collider correspond to scenarios where the total luminosities are assumed to be 5 and 20 fb$^{-1}$. The last row also includes the derived precision on the total Higgs width from the fits. 
\label{t:Kappa_sensitivity_2}    }
\end{table}

%======================================================

\subsection{The Standard Model Effective Field Theory}

The Standard Model Effective Field Theory (SMEFT) framework provides a systematic parameterization of the effects of new physics, with the assumption that the electroweak symmetry breaking is linearly realized and the new physics is significantly heavier than the electroweak scale~\cite{Buchmuller:1985jz, Grzadkowski:2010es}. Upon truncation of the effective theory expansion at a given order, based on the EFT power counting and the experimental precision, the general bottom-up approach is to simultaneously fit all possible measurements to all the EFT operator coefficients contributing to the corresponding processes on a non-redundant basis. We restrict our analysis to
effects from dimension-six operators and focus our attention to those that can modify the Higgs boson processes.
To extract most of the Higgs boson properties, at least at the leading order, it is sufficient to consider the Higgs and electroweak boson measurements in the fits. This has been done for measurements at both the LHC and LEP/SLD ~\cite{Falkowski:2015jaa, Ellis:2018gqa, Dawson:2020oco, Ellis:2020unq, Ethier:2021bye, Almeida:2021asy} and the future $e^+e^-$ colliders~\cite{Barklow:2017suo, deBlas:2019rxi, DeBlas:2019qco}.  The electroweak measurements are essential in the analysis since a number of operators contribute to both Higgs and electroweak processes, and the electroweak measurements provide important constraints on them. This includes the $Z$-pole observables and the $e^+e^-(\mu^+\mu^-)\to W^+W^-$ process at lepton colliders, as well as the diboson ($pp\to WW/WZ$) processes at the LHC. Without these constraints, flat directions may appear in the space of SMEFT interactions contributing to Higgs processes, which could significantly reduce the overall sensitivity in a global framework.  

Compared with a future $e^+e^-$ collider, the 125\,GeV muon collider lacks both the $Z$-pole measurements and the one of $\mu^+\mu^- \to W^+W^-$ (which requires $\sqrt{s} >160$\,GeV).  This disadvantage can, however, be overcome in several ways. First, the electroweak measurements at the LEP/SLD and HL-LHC 
already provide reasonable constraints on the corresponding operator coefficients. Second, in a plausible scenario, a future $e^+e^-$ collider would further improve these measurements, providing important complementarity to the muon-collider measurements. Third, perhaps most importantly, the high energy runs of a muon collider would be able to probe many of these operators with unprecedented sensitivities due to their large energy enhancement~\cite{Buttazzo:2020uzc}.  
This also includes operators that are conventionally thought to be better probed by the $Z$-pole measurements. For instance, operators such as $\mathcal{O}_{H\ell} = i  H^\dagger \overleftrightarrow{D_\mu} H \bar{\ell}_L \gamma^\mu \ell_L$ modify the $Z\ell^+\ell^-$ coupling, but also contribute to the process $\ell^+\ell^- \to h Z$ via a $h Z\ell^+\ell^-$ contact interaction, and their effects thus grow with energy ($\propto E^2$). 
We will consider the impact of these high energy measurements in our analysis.  

The following measurement inputs are used in our analysis:
We use the same inputs for the Higgs measurements as in the $\kappa$ analysis. We implement the $Z$-pole measurements at LEP and SLD in Ref.~\cite{ALEPH:2005ab}. For the $Z$-pole measurements at a future $e^+e^-$ collider, we use the CEPC inputs in Ref.~\cite{CEPCStudyGroup:2018ghi}. We implement the HL-LHC diboson analysis results in Ref.~\cite{Grojean:2018dqj}. For the $W^+W^-$ analysis at a $e^+e^-$ collider, we follow Ref.~\cite{DeBlas:2019qco} and obtain the results from an optimal observable analysis~\cite{Diehl:1993br}. We also perform a similar optimal observable analysis to the measurements of $\mu^+\mu^-\to W^+W^-$ as well as the $\mu^+\mu^-\to hZ$ process at the high-energy runs of the muon collider. While the $e^+e^- \to W^+W^-$ process was also measured at LEP 2~\cite{ALEPH:2013dgf}, the measurement precision was relatively low, and they do not make a significant impact once the HL-LHC diboson measurements are included. For simplicity, they are not included in our analysis.  

\begin{figure}
\centering
\includegraphics[width=\linewidth]{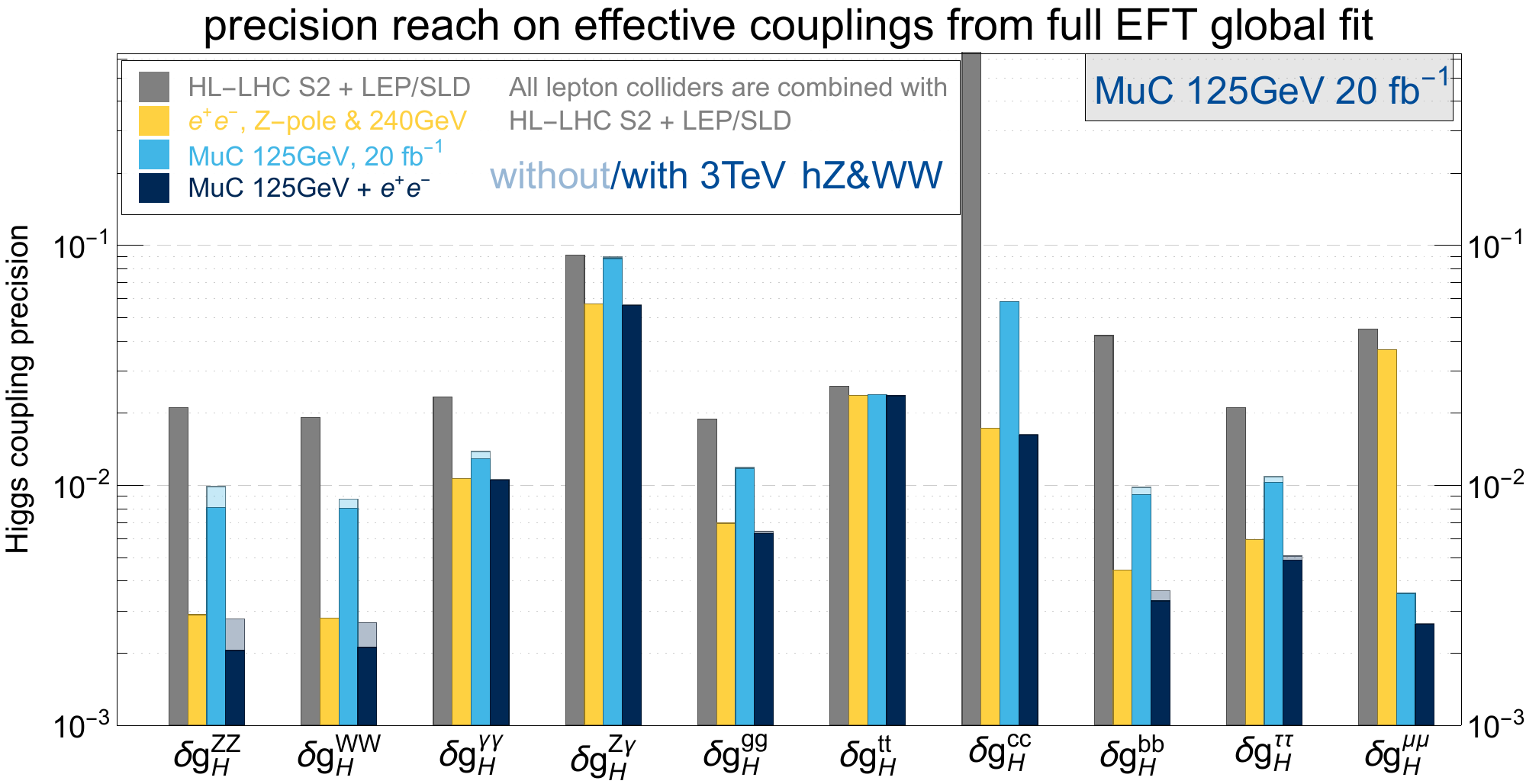}
\caption{
The one-sigma precision reach on the effective Higgs couplings 
from a global fit of the Higgs and electroweak measurements in the SMEFT framework. The four columns represent the HL-LHC S2 scenario with electroweak measurements at LEP and SLD, a circular $e^+e^-$ collider with center-of-mass energy up to $240\,$GeV, a muon collider at 125\,GeV with a total integrated luminosity of $20\ifb$, and the combination of the $e^+e^-$ and the muon collider, respectively. 
The measurements are combined with the HL-LHC S2 and LEP/SLD measurements for all the lepton collider scenarios. For the last two scenarios, the diboson $hZ$ and $WW$ measurements at a 3\,TeV muon collider ($1\iab$) are also considered to show their impact on different operators.\footnote{Note that we do not include the Higgs precision input from a 3 TeV muon collider here. One can research on the complementarity between 125 GeV muon collider with high energy muon collider in a future study.} Results with (without) the 3\,TeV $hZ/WW$ measurements are shown with solid (light-shaded) columns. 
}
\label{fig:eft1}
\end{figure}
\begin{figure}
\centering
\includegraphics[width=\linewidth]{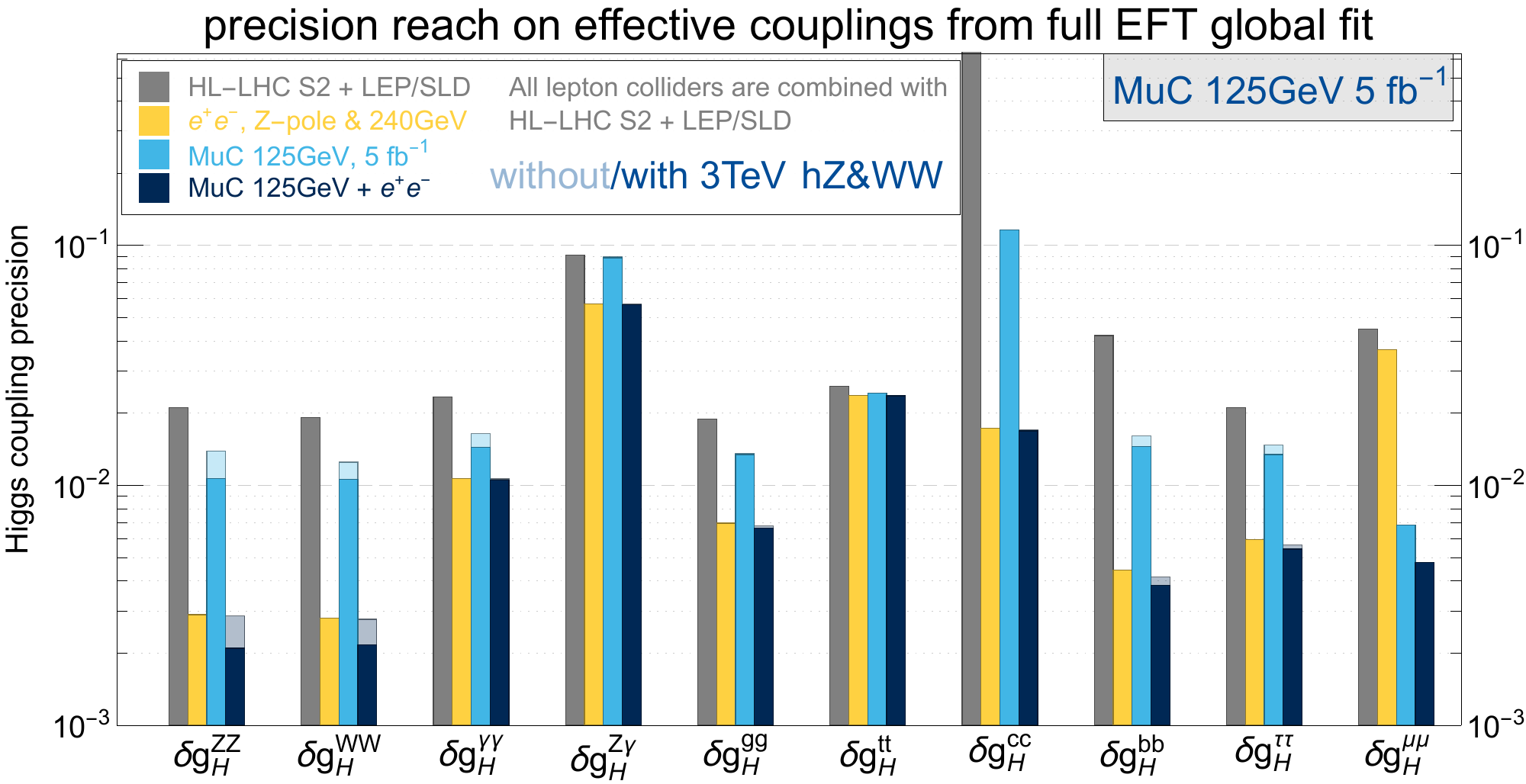}
\caption{
Same as \autoref{fig:eft1} but assuming $5\ifb$ at the 125\,GeV muon collider.
}
\label{fig:eft2}
\end{figure}

 Following Ref.~\cite{DeBlas:2019qco} (see also Ref.~\cite{deBlas:2019rxi}), we perform a global fit to the Higgs and electroweak measurements with a total number of 28 independent CP-even Wilson coefficients. Since our focus is on the Higgs measurements and the corresponding Higgs coupling constraints, we present the result of the fit in terms of effective  Higgs couplings~\cite{Barklow:2017suo, deBlas:2019rxi, DeBlas:2019qco}, while treating all other couplings/coefficients as nuisances parameters. 
The result is shown in \autoref{fig:eft1} in terms of the one-sigma precision of these effective couplings, denoted as $\delta g^{XX}_H$. 
The numerical results of our analysis are provided in \autoref{t:eft}.   
Four scenarios are considered, which are the HL-LHC S2 combined with the electroweak measurements at LEP and SLD, a circular $e^+e^-$ collider with center-of-mass energy up to 240\,GeV, a 125\,GeV muon collider with 20$\ifb$ integrated luminosity, and the combination of the $e^+e^-$ and the muon collider. The HL-LHC S2 and LEP/SLD measurements are also included in the fit for the last three scenarios. For the muon collider, we consider two cases: the 125\,GeV run only, and the $hZ$ and $WW$ measurements at 3 TeV (assuming a total luminosity of $1\iab$), as mentioned earlier. They are shown by the light-shaded and solid columns, respectively.
We also consider the scenario of a $125$\,GeV muon collider with a total integrated luminosity of $5\fbi$, the results of which are shown in \autoref{fig:eft2}. The table includes the results of both scenarios.  

Several important observations can be made when comparing the muon collider results with the ones of other collider scenarios.  
For $\delta g^{\mu\mu}_H$, as expected, the 125\,GeV muon collider provides an unprecedented sensitivity, which improves the precision reach of the  HL-LHC and a $e^+e^-$ collider by at least one order of magnitude. For most of the couplings (except $\delta g^{Z\gamma}_H$ and $\delta g^{tt}_H$), the 125\,GeV muon collider provides a significant improvement to the HL-LHC results. This is even true for $\delta g^{\gamma\gamma}_H$. 
While the $h\to\gamma\gamma$ channel is poorly measured at the muon collider, the improvements on the other couplings and the total Higgs width lift certain flat directions associated with  $\delta g^{\gamma\gamma}_H$ in the LHC measurements and improves its reach in the global analysis. While a $e^+e^-$ collider generally provides better constraints on most Higgs couplings, the 125\,GeV muon collider still offers important complementarity by providing a different production channel. In particular, $\delta g^{b b}_H$ receives a non-negligible improvement from the combination of the two collider scenarios. 
Most of the observed patterns are similar to the results of the $\kappa$-fit. However, a major difference can be found for the $hZZ$ and $hWW$ couplings. While they are treated as free parameters in the $\kappa$-fit, in the SMEFT framework, we always have $\delta g^{ZZ}_H \approx \delta g^{WW}_H$ due to the constraints on custodial-symmetry-violating effects from electroweak measurements. Their difference is relatively larger when the electroweak precision constraints are weaker, for instance, in a 125\,GeV muon collider alone without the $e^+e^-$ collider or the high energy electroweak measurements.

\begin{table}[t]
  \centering
     \vspace{0.25cm}
    {\small
      \begin{tabular}{c|c|c|cc|cc|cc|cc} 
        \ctoprule
        %\crowcolor
        &
        HLLHC &
        HLLHC &
        \multicolumn{4}{c|}{HL-LHC+ $\mu$-coll.} &
        \multicolumn{4}{c}{HL-LHC+ $\mu$-coll.\&$e^+e^-$}   \\
        Coupling &
         &
         
        + $e^+e^-$  &
        
        \multicolumn{2}{c|}{w/o 3\,TeV EW} &
        \multicolumn{2}{c|}{w 3\,TeV EW} &

        \multicolumn{2}{c|}{w/o 3\,TeV EW} &
        \multicolumn{2}{c}{w 3\,TeV EW}  \\
        &&&5 fb$^{-1}$&20 fb$^{-1}$&5 fb$^{-1}$&20 fb$^{-1}$&5 fb$^{-1}$&20 fb$^{-1}$&5 fb$^{-1}$&20 fb$^{-1}$\\
        
        \cmrule
        &&&&&&&&&\\ [-0.25cm]
%-----------------------------------
        $\delta g^{ZZ}_H$ &
        2.1 & 0.29 & 1.4 & 0.99 & 1.1 & 0.81 & 0.29 & 0.28 & 0.21 & 0.20 \\ [0.1cm]
          
        $\delta g^{WW}_H$ &
         1.9 & 0.28 & 1.2 & 0.88 & 1.1 & 0.80 & 0.28 & 0.27 & 0.22 & 0.21 \\ [0.1cm]
          
        $\delta g^{\gamma\gamma}_H$ &
         2.3 & 1.1 & 1.6 & 1.4 & 1.4 & 1.3 & 1.1 & 1.1 & 1.1 & 1.0 \\ [0.1cm]
          
        $\delta g^{Z\gamma}_H$ &
          9.1 & 5.7 & 9.0 & 8.9 & 8.9 & 8.8 & 5.7 & 5.7 & 5.7 & 5.6 \\ [0.1cm]
          
        $\delta g^{gg}_H$ &
         1.9 & 0.70 & 1.4 & 1.2 & 1.3 & 1.2 & 0.68 & 0.64 & 0.66 & 0.63 \\ [0.1cm]
          
        $\delta g^{tt}_H$ &
         2.6 & 2.4 & 2.4 & 2.4 & 2.4 & 2.4 & 2.4 & 2.4 & 2.4 & 2.4 \\ [0.1cm]  
          
        $\delta g^{cc}_H$ &
        -- & 1.7 & 12. & 5.8 & 12. & 5.8 & 1.7 & 1.6 & 1.7 & 1.6 \\ [0.1cm]    
          
        $\delta g^{bb}_H$ &
        4.2 & 0.44 & 1.6 & 0.98 & 1.4 & 0.92 & 0.41 & 0.36 & 0.38 & 0.33 \\ [0.1cm]    
          
        $\delta g^{\tau\tau}_H$ &
        2.1 & 0.59 & 1.5 & 1.1 & 1.3 & 1.0 & 0.56 & 0.51 & 0.54 & 0.49 \\ [0.1cm] 
          
        $\delta g^{\mu\mu}_H$ &
        4.5 & 3.7 & 0.68 & 0.35 & 0.68 & 0.35 & 0.47 & 0.26 & 0.47 & 0.26 \\ [0.1cm] 
          
        \cbottomrule
      \end{tabular}
    }
\caption{Results from the SMEFT fit in terms of the effective Higgs couplings.  ``3\,TeV EW'' denotes the measurements of $\mu^+\mu^- \to hZ$ and $\mu^+\mu^- \to W^+W^-$ at a 3\,TeV muon collider ($1\abi$), implemented using an optimal observable analysis.  
    \label{t:eft}}
\end{table}

The interplay with the high energy runs of the muon collider is also an important aspect of the SMEFT global fit. First of all, the high energy runs provide a huge Higgs sample from the $WW$ fusion process, which could improve the Higgs coupling precision by a significant amount. Careful treatments on the backgrounds and systematic uncertainties are needed to obtain the projected precision reaches of these measurements. We leave such an analysis for future studies. On the other hand, hard processes such as $\mu^+\mu^- \to hZ$ or $\mu^+\mu^- \to WW$ have smaller cross-sections at higher energies, but they provide much stronger constraints on the operators that exhibit energy enhancement, %.  
which could potentially make their uncertainties negligible in the Higgs coupling fit~\cite{Buttazzo:2020uzc}. However, flat directions may still be present among the operator coefficients in the global analysis. To take account of their effects, we implement an optimal observable analysis for $hZ$ and $WW$ instead of directly using the results of Ref.~\cite{Buttazzo:2020uzc}. Our result indeed shows that these high energy measurements provide a non-negligible improvement on the Higgs couplings constraints of the $125\,$GeV run, especially for $\delta g^{ZZ}_H$ and $\delta g^{WW}_H$.  
We also notice a strong complementarity between the Higgs measurements at a $e^+e^-$ collider and the $hZ/WW$ measurements at a high energy muon collider, as the combined results (shown in the last column) also receives significant improvements on $\delta g^{ZZ}_H$ and $\delta g^{WW}_H$ with the inclusion of the $hZ/WW$ measurements.  
We have also checked that the inclusion of $hZ/WW$ measurements of the 10 and 30\,TeV run does not significantly change the results, as the corresponding directions have already been constrained sufficiently well by the 3\,TeV measurements, relative to the Higgs couplings.

%\clearpage

\section{Conclusion}
\label{sec:conclusion}

In this paper, we have presented the results of an optimized framework to determine the Higgs parameters from a lineshape scan and fit at a 125 GeV muon collider. Many subtle considerations of the scan and fitting framework were explored here, including proper scan range, minimal scan steps, scan luminosity distribution, effective cross-sections, correlations between fitting parameters, and others. The important effects, such as BES and ISR, are consistently considered, and the fits are performed with a large simulated lineshape data sample. We propose to conduct resonant lineshape searches in the $m_H\pm 8~\mev$ window. The scan features 11 evenly spaced energy incremental steps and an even distribution of beam luminosity. Such a proposal reasonably optimizes the precision of extraction of the Higgs lineshape parameters and leaves room for potential systematics from various sources.

From the lineshape fit, the width determination from the threshold scan is dominantly coming from the $b\bar b$ and $WW^*$ channels. The signal to background ratio is approximately 1:2 and 40:1 for $b\bar b$ and $WW^*$, after considering different signal modes, background, and detection efficiencies. Although the $b\bar b$ channel has more than three times the statistics of the $WW^*$ channels, their contribution to the Higgs width determination is comparable due to different sources of background. We also discuss the relation of the lineshape scan fit and individual exclusive channel measurements.

All major Higgs decay channels are studied in our analysis under the consistent scan framework. After considering the BES, ISR, and off-shell suppression during the scan process, we can evaluate the effective Higgs cross-sections and study how various channels stand out from the background. The effective Higgs cross-sections and the precision result of individual channels are tabulated in \autoref{tab:sensitivity}. We report the Higgs decay channels of $b\bar b$, $c\overline c$, and $gg$, and also take into account the cross-correlations between these channels and the angular distribution information. We studied the $\tau^+\tau^-$ decay in various modes and the $\gamma\gamma$ channel with angular correlations and experimental cuts. The Higgs decays into $VV^*$ channels yield many different possible final states, including all hadronic, semi-leptonic, fully leptonic, with and without missing energy, and some of which can be mediated by both $WW^*$ and $ZZ^*$ intermediate states with interference, e.g., for the $4j$ or the $\ell^+ \ell^- \nu\bar{\nu}$ final states. We study them individually and report their corresponding precision. 

These new studies on the Higgs lineshape and exclusive channels enable us to clarify the Higgs physics potential of a 125 GeV muon collider and put this option into a more global picture. For this purpose, we performed global fits, including the Higgs projections studied here, both in the $\kappa$ framework and in the dimension-six SMEFT. We consider several scenarios, showing the resonant muon collider Higgs factory results in combination with the HL-LHC and with potential future $e^+e^-$ Higgs factories. These results are summarized in \autoref{fig:kapparesult}, \autoref{fig:kapparesult5ifb}, \autoref{fig:eft1}, and well detailed in \autoref{t:Kappa_sensitivity}, \autoref{t:Kappa_sensitivity_2} and \autoref{t:eft}. We show these various scenarios for the resonant muon collider Higgs factory for both $5~\fbi$ and $20~\fbi$ of integrated luminosities. 

To summarize, the Higgs physics program at a 125 GeV muon collider can improve the HL-LHC determination of the Higgs coupling by a factor of a few for most couplings. It provides independent and complementary information about the Higgs boson compared to $e^+e^-$ Higgs factories. 
One could explore the physics potential in many new directions, such as further optimizing the lineshape scan, the prescan, understanding the systematics, and robustness against various technological pathways. In particular, one can study the synergies between the resonant muon collider Higgs factory with its high-energy runs. 
The critical question, how well such a machine fits into the ambitious roadmap of multi-(even tens of) TeV muon colliders, remains open to critical development on the accelerator end. The value of a resonant muon collider Higgs factory also varies as the global high energy physics future collider program evolves. Our work provides a clear and solid ground in understanding the physics reach, which provides a clean foundation for further physics discussions.

\acknowledgements{
The authors would like to thank Mario Greco, Tao Han, Patrick Meade, Mark Palmer, Sergo Rigiriolo and Hannsjoerg Weber for helpful discussions. 
The work of J.B. has been supported by the FEDER/Junta de Andaluc\'ia project grant P18-FRJ-3735. J.G. is supported by National Natural Science Foundation of China (NSFC) under grant No.~12035008.  Z.L. would like thank the organizers of the PITT-PACC muon collider workshop, in the preparation of the overview talk, this work was stimulated. Z.L. is supported in part by the U.S. Department of Energy (DOE) under grant No. DE-SC0022345. Z.L. would like to thank Aspen Center for Physics, supported by National Science Foundation (NSF) grant PHY-1607611, where part of this work was completed. Z.L. acknowledges the Minnesota Supercomputing Institute (MSI) at the University of Minnesota for providing resources that contributed to the research results reported within this paper \href{http://www.msi.umn.edu}{http://www.msi.umn.edu}.
}

%\newpage

\appendix

\bibliographystyle{utphys}%{utphys}%{plainnat}
\bibliography{reference}

\end{document}